\pdfoutput=1
\documentclass[preprint,floats,floatfix,aps,epsfig,nofootinbib,letter]{revtex4}
\usepackage{graphics,graphicx,psfrag}
\usepackage{amsmath,amssymb}
\usepackage[sort&compress]{natbib}
\usepackage{cancel}
\usepackage{bm}
\usepackage{verbatim}
\usepackage{multirow}
\usepackage{subfigure}
\usepackage{ulem}
\usepackage{color}
\usepackage[linktocpage=true,bookmarksnumbered=true,colorlinks=true,plainpages,linkcolor=blue,citecolor=red]{hyperref}
\def\beq{\begin{equation}}
\def\eeq{\end{equation}}
\def\bea{\begin{eqnarray}}
\def\eea{\end{eqnarray}}
\def\nn{\nonumber}

\def\ifb{\rm fb^{-1}}

\def\tev{\rm TeV}
\def\zp{Z^\prime}
\def\dcol{D_{\rm{col}}}
\def\mres{M_{\rm{res}}}

\begin{document}
%

\preprint{
{\vbox {
\hbox{\bf MSUHEP-130619}
}}}
\vspace*{2cm}

\title{Distinguishing Color-Octet and Color-Singlet Resonances at the Large Hadron Collider\\} 
\vspace*{0.25in}   
\author{{Anupama Atre, R. Sekhar Chivukula, Pawin Ittisamai \\ 
and Elizabeth H. Simmons}\footnote{avatre@pa.msu.edu\\sekhar@pa.msu.edu\\ ittisama@msu.edu\\esimmons@pa.msu.edu} }  
\affiliation{\vspace*{0.1in}
Department of Physics and Astronomy\\
Michigan State University, East Lansing U.S.A.\\}
\vspace*{0.25 in} 


\begin{abstract}
\vspace{0.5cm}
\noindent

Di-jet resonance searches are simple, yet powerful and model-independent, probes for discovering new particles at hadron colliders. Once such a resonance has been discovered it is important to determine the mass, spin, couplings, chiral behavior and color properties to determine the underlying theoretical structure. We propose a new variable which, in the absence of decays of the resonance into new non-standard states, distinguishes between color-octet and color-singlet resonances. To keep our study widely applicable we study phenomenological models of color-octet and color-singlet resonances in flavor universal as well as flavor non-universal scenarios. We present our analysis for a wide range of mass ($2.5 - 6\ \tev$), couplings and flavor scenarios for the LHC with center of mass energy of $14\ \tev$ and varying integrated luminosities of $30,\ 100,\ 300\ \rm{and}\ 1000\ \ifb$. We find encouraging results to distinguish color-octet and color-singlet resonances for different flavor scenarios at the LHC. 

\end{abstract}

\maketitle

\section{Introduction}
\label{sec:intro}

Hadron colliders are a rich source for the production of new resonances with strong coupling due to colored particles in the initial state. A particularly simple and powerful probe of new colored resonances is the di-jet channel where the resonance decays to two partons. Each successive hadron collider with an increase in center of mass energy and integrated luminosity has been able to probe di-jet resonances with higher masses. Many well motivated theories of physics beyond the Standard Model (SM) predict new particles that give rise to signatures in the di-jet channel. These new particles can have different spin and color structure and a sample of such possibilities is listed below. 

The color-octet vector boson arises as a result of extending the gauge group of the strong sector. The chiral structure of the couplings between quarks and the color-octet varies and the couplings can either be  flavor universal or flavor non-universal. Examples of flavor universal scenarios are the axigluon \cite{Frampton:1987dn, Frampton:1987ut} and coloron \cite{Chivukula:1996yr, Simmons:1996fz} where all quarks are charged under the same $SU(3)$ group. Flavor non-universal scenarios appear in the case of the topgluon where the third generation quarks are assigned to one $SU(3)$ group and the light quarks to the other \cite{Hill:1991at, Hill:1994hp} and the axigluon where different chiralities of the same quark can be charged under different groups \cite{Antunano:2007da, Ferrario:2009bz, Frampton:2009rk, Rodrigo:2010gm, Chivukula:2010fk, Tavares:2011zg} . Other examples include Kaluza-Klein (KK) gluons which are excited gluons in extra-dimensional models \cite{Dicus:2000hm}, technirhos which are composite colored vector mesons found in technicolor~\cite{Farhi:1980xs,Hill:2002ap, Lane:2002sm}, models that include colored technifermions and low-scale string resonances ~\cite{Antoniadis:1990ew}. 

The electrically neutral color-singlet vector boson, collectively called a $Z^\prime$, also appears in many beyond SM physics scenarios and can originiate from extending the electroweak $U(1)$ or $SU(2)$ gauge group. The $Z^\prime$ can also have flavor universal \cite{Senjanovic:1975rk, Georgi:1989ic, Georgi:1989xz} or flavor non-universal couplings to fermions \cite{Muller:1996dj, Malkawi:1996fs, Chivukula:1994mn}. For reviews of $Z^\prime$ models, see Refs.~\cite{Langacker:2008yv, Leike:1998wr, Hewett:1988xc} and the references therein. Other examples of color-singlet states probed in the di-jet channel include the spin-2 gravitons in Randall-Sundrum models~\cite{Randall:1999ee,Randall:1999vf} which are KK excitations of the gravitational field~\cite{Davoudiasl:1999jd, Bijnens:2001gh} and the electrically charged color-singlet vector boson, namely the $W^\prime$~\cite{Eichten:1984eu}. 

Searches for new resonances in the di-jet channel have been performed at the CERN $\rm{S\bar{p}pS}$~\cite{Arnison:1986vk, Alitti:1993pn}, Tevatron~\cite{Abe:1989gz, Abe:1995jz, Abe:1997hm, Abazov:2003tj, Aaltonen:2008dn} and the Large Hadron Collider (LHC)~\cite{Aad:2010ae, Khachatryan:2010jd, Chatrchyan:2011ns, Aad:2011fq, atlas:2012nma,ATLAS:2012pu,ATLAS:2012qjz,Chatrchyan:2013qha, CMS:kxa}. Once a resonance has been discovered the next step is to measure the properties of the resonance. The di-jet invariant mass $m_{jj}$ and the angular distributions of energetic jets relative to the beam axis are sensitive observables to determine the mass and spin of the resonance. The coupling strength of the resonance to SM quarks can be constrained using the cross section of the di-jet channel and the chiral structure can be determined by combining the di-jet channel with the channel where the resonance is produced in association with a SM electroweak gauge boson~\cite{Atre:2012gj}. As a next step, we explore the question of determining the color structure of the resonance produced in the di-jet channel; in particular we explore whether the resonance is a color-octet or a color-singlet state. 

To distinguish a color-octet and a color-singlet resonance produced in the di-jet channel we introduce a new variable called the color discriminant variable. Assuming that the new resonance decays only to quarks, this variable reflects the color structure of the resonance and is constructed from measurements which can be made in the di-jet discovery channel, namely the di-jet cross section, mass and width of the resonance. We demonstrate the utility of this variable in distinguishing color-octet and color-singlet resonances using the simple flavor universal example of a coloron and a leptophobic $\zp$. We also demonstrate the robustness of this method using more general cases like flavor non-universal scenarios. We study the sensitivity of the LHC with center of mass (c.m.) energy of $14\ \tev$ and integrated luminosities of $30, 100, 300\  \rm and\  1000\  \rm \ifb$ to distinguish color-octet and color-singlet resonances. Motivated by current constraints on di-jet resonances and future prospects for discovery, we probe masses ranging from $2.5 - 6\ \tev$ with various couplings and widths. 

The rest of the paper is organized as follows. In Sec.~\ref{sec:genpar} we describe phenomenological models for a coloron and a leptophobic $Z^\prime$ used as illustrative examples in this article.  We introduce a color discriminant variable to distinguish color-octet and color-singlet resonances in Sec.~\ref{sec:coldis}. We discuss the current constraints from collider searches and the allowed regions in parameter space in Sec.~\ref{sec:constr}  and the uncertainties involved in measuring the color discriminant variable at the LHC in Sec.~\ref{sec:sens}. We present our results in Sec.~\ref{sec:reslt} and conclusions in Sec.~\ref{sec:summ}. A discussion about uncertainties relevant to our analysis is presented in Appendix~\ref{sec:app}. 

\section{General Parameterization}
\label{sec:genpar}

Color-octet and color-singlet resonances of interest to our study may be motivated in many beyond the SM physics scenarios as described in the introduction. Therefore we study a phenomenological model of color-octet and color-singlet resonances independent of the underlying theory to keep our study widely applicable. We will assume that there are no additional colored states into which the resonance can decay. If new light states are present, studying the properties of the decays of the coloron or $\zp$ into these new particles will be instructive (see Refs.~\cite{Dobrescu:2007yp, Chivukula:2013hga} and references therein). The couplings of the color-octet and color-singlet resonances to SM quarks can all be the same in the simple flavor universal scenario and can all be independent in the most general flavor non-universal case. In this section, we present details about the parameterization of the interactions of color-octet and color-singlet resonances for two cases, the flavor universal scenario and an illustrative flavor non-universal scenario.  

The interaction of a color-octet resonance $C_\mu$ with the SM quarks $q_i$ has the form
\beq
\mathcal{L}_C  = i g_s\bar{q}_i\gamma^\mu\left( g_{C_L}^i P_L + g_{C_R}^i P_R \right) q_i C_\mu, \\
\label{eq:colcoupl}
\eeq
where $C_\mu = C_\mu^a t^a$ with $t^a$ an $ \text{SU}(3) $ generator, $g_{C_L}^i$ and $g_{C_R}^i$ denote left and right chiral coupling strengths of the color-octet to the SM quarks relative to the QCD coupling $g_s$, the projection operators have the form $P_{L,R} = (1 \mp \gamma_5)/2$ and the quark flavors run over $ i=u,d,c,s,t,b.$ We will denote the color-octet resonance by $C$ and its chiral couplings to light quarks by $g_{C_{L,R}}^q$ and to the third generation by $g_{C_{L,R}}^t$ and use the terms color-octet and coloron interchangeably. 

The color-octet resonance with the interactions as in Eq.~(\ref{eq:colcoupl}) decays primarily to two jets or a top pair and its decay width is given by
\bea
\nn
\Gamma_C &=& \alpha_s \frac{M_C}{12} \Big[ 4 \left(g_{C_L}^{q\ 2} + g_{C_R}^{q\ 2}\right) + \left( g_{C_L}^{t\ 2} + g_{C_R}^{t\ 2} \right) \\
&+& \left[ (g_{C_L}^{t\ 2} + g_{C_R}^{t\ 2}) (1 -  \mu_t)  + 6g_{C_L}^{t}g_{C_R}^{t} \mu_t \right] \sqrt{1 - 4 \mu_t} \Big],     
\label{eq:colwidfull}
\eea
where $M_C$ and $\Gamma_C$ are the mass and intrinsic width of the color-octet respectively. Decays to top quarks are modified by the kinematic factors involving $\mu_{t} = m_{t}^2/M_C^2$ with $m_{t}$ the top quark mass. Strictly speaking, the bottom quark's contribution to the width is modified by factors involving $\mu_b=m_b^2/M_C^2$ but we ignore these factors since $m_b^2 << M_C^2$. For a color-octet that is heavy compared to the top quark, the expression for the total width of a color-octet in Eq.~(\ref{eq:colwidfull}) simplifies to
\beq
\Gamma_C = \alpha_s \frac{M_C}{12} \Big[ 4 \left(g_{C_L}^{q\ 2} + g_{C_R}^{q\ 2}\right) + 2 \left( g_{C_L}^{t\ 2} + g_{C_R}^{t\ 2} \right) \Big].     
\label{eq:colwidnom}
\eeq

We parameterize the interaction of a color-singlet similarly to that of the color-octet. If the color-singlet has tree level couplings to SM leptons also, it can be easily distinguished from a color-octet as the octet has no decays to leptons. Hence we consider only the leptophobic variant of the color-singlet, as this can mimic a color-octet resonance in the di-jet channel. We will call such a resonance a leptophobic $Z^\prime$ henceforth. The interactions of a leptophobic $\zp$ with the SM quarks are given by
\beq
\mathcal{L}_{\zp}  = i g_w\bar{q}_i \gamma^\mu\left( g_{\zp_L}^i P_L + g_{\zp_R}^i P_R \right) q_i \zp_\mu, 
\label{eq:zpcoupl}
\eeq
where $g_{\zp_L}^i$ and $g_{\zp_R}^i$ denote left and right chiral coupling strengths of the leptophobic $\zp$ to the SM quarks relative to the weak coupling $g_w = e/\sin\theta_W$ and the quark flavors run over $ i=u,d,c,s,t,b.$ Again the chiral couplings of the $\zp$ to light quarks are denoted by $g_{\zp_{L,R}}^q$ and to the third generation by $g_{\zp_{L,R}}^t$. 

The leptophobic $\zp$ with the interactions as in Eq.~(\ref{eq:zpcoupl}) decays primarily to two jets or a top pair and its decay width is given by
\bea
\nn
\Gamma_{\zp} &=& \alpha_w \frac{M_{\zp}}{2} \Big[ 4 \left(g_{\zp_L}^{q\ 2} + g_{\zp_R}^{q\ 2}\right) + \left( g_{\zp_L}^{t\ 2} + g_{\zp_R}^{t\ 2} \right) \\
&+& \left[ (g_{\zp_L}^{t\ 2} + g_{\zp_R}^{t\ 2}) (1 -  \mu_t)  + 6g_{\zp_L}^{t}g_{\zp_R}^{t} \mu_t \right] \sqrt{1 - 4 \mu_t} \Big],     
\label{eq:zpwidfull}
\eea
where $M_{\zp}$ and $\Gamma_{\zp}$ are the mass and intrinsic width of the leptophobic $\zp$ respectively. Similar to the case of the color-octet, we can neglect the kinematic factors involving the mass of the top and bottom quarks when the $\zp$ is much heavier than the SM quarks. Hence the expression for the decay width of a $\zp$ as given in Eq.~(\ref{eq:zpwidfull}) simplifies to
\beq
\Gamma_{\zp} = \alpha_w \frac{M_{\zp}}{2} \Big[ 4 \left(g_{\zp_L}^{q\ 2} + g_{\zp_R}^{q\ 2}\right) + 2 \left( g_{\zp_L}^{t\ 2} + g_{\zp_R}^{t\ 2} \right)\Big]. 
\label{eq:zpwidnom}
\eeq
In the rest of the article we will only consider color-octet and color-singlet resonances that are much heavier than the top quark. 

\subsection{Flavor Universal Scenario}
\label{sec:flavuniv}

In the flavor universal scenario all SM quarks have the same coupling to the color-octet resonance, {\it i.e.} $g_{C_{L,R}}^q = g_{C_{L,R}}^t = g_{C_{L,R}}$. The expression for the decay width of a color-octet as given in Eq.~(\ref{eq:colwidnom}) simplifies further to
\beq
\Gamma_C = \frac{\alpha_s}{2} M_C \left( g_{C_L}^2 + g_{C_R}^2\right),  
\label{eq:colwiduniv}
\eeq
and the branching fraction for the color-octet resonance to decay to jets obeys the simple relation
\beq
BR( C \rightarrow jj ) = 5/6, \hspace{1cm} {\mbox {\text where}\  j=u,d,c,s,b}. 
\label{eq:axibruniv}
\eeq

Similarly, the decay width of a $\zp$ with flavor universal couplings to SM quarks simplifies to
\beq
\Gamma_\zp = 3 \alpha_w M_\zp \left( g_{\zp_L}^2 + g_{\zp_R}^2\right),
\label{eq:zpwiduniv}
\eeq
where $\alpha_w = g_w^2/4\pi$ and $g_{\zp_{L,R}}^q = g_{\zp_{L,R}}^t = g_{\zp_{L,R}}$. The branching fraction for a $\zp$ to decay to jets obeys the simple relation
\beq
BR( \zp \rightarrow jj) = 5/6, \hspace{1cm} {\mbox {\text where}\  j=u,d,c,s,b}
\eeq

Note that although the width of the coloron is proportional to the strong coupling 
($\alpha_{\rm s}(m_Z) \simeq 0.12$) and that of the leptophobic $\zp$ is proportional to the weak coupling ($\alpha_{\rm w} \simeq 0.04$), the two resonances will have comparable widths when the couplings $g_{L}^2 + g_{R}^2$ are the same. This is due to the difference in the color factors for the two resonances. 

\subsection{An Illustrative Flavor Non-universal Scenario}
\label{sec:flavnonuniv}

The couplings of a color-octet and color-singlet resonance to SM quarks can all be independent in the most general flavor non-universal scenario. While it is desirable to study the most general case, it is computationally cumbersome and beyond the scope of this study. Instead we consider an intermediate scenario where the couplings of the color-octet and the color-singlet to quarks in the third generation are different from the couplings of the quarks in the first two generations. An interesting example of such a scenario is that of a color-octet resonance (described in Refs.~\cite{Antunano:2007da, Frampton:2009rk, Ferrario:2009bz, Rodrigo:2010gm}) that can enhance top-pair forward-backward asymmetry observed at the Tevatron~\cite{Abazov:2007ab, Aaltonen:2008hc, Abazov:2011rq, Aaltonen:2011kc, CDF:2012xba}. For the $Z^\prime$, models related to strong dynamics typically feature a $Z^\prime$ that couples with an enhanced strength to quarks in the third family (see Ref.~\cite{Langacker:2008yv} for a review of such cases).

The couplings of a color-octet or color-singlet resonance to SM quarks in the flavor non-universal scenario we consider can be parametrized by
\beq
g_{C/\zp_{L,R}}^{t} \equiv \xi g_{C/\zp_{L,R}}^{q}\ ,
\label{eq:xi}
\eeq
where $t=t,b$ and $q=u,d,c,s$. The change in the total decay width of a color-octet and color-singlet in the flavor non-universal scenario compared to the width in the flavor-universal case (given by Eq.~(\ref{eq:colwiduniv}) and Eq.~(\ref{eq:zpwiduniv}) respectively) is given by
\beq
\Gamma^{\rm{non-universal}} =\Gamma^{\rm{universal}}  \left( \frac{4  + 2 \xi^2}{6} \right).
\label{eq:widnonuniv}
\eeq
The branching fraction of the color-octet and color-singlet to jets, where the jets are defined to include $j=u,d,c,s,b$ changes from 
\beq
Br(C/\zp \to jj) = 5/6 
\eeq
in the flavor universal case to 
\beq
Br(C/\zp \to jj) = \frac{4 + \xi^2}{4 + 2\xi^2}\, 
\label{eq:brnonuniv}
\eeq
in the flavor non-universal case. Hence the branching fraction to jets in the flavor non-universal case decreases to $1/2$ from $5/6$ in the flavor universal case as $\xi$ becomes large. This tendency towards smaller branching fraction in the flavor non-universal case tends to decrease the overall di-jet cross section for the resonance.  The production rate stays the same, as it is dominated by the contribution from the first (and to a small extent from the second) generation quarks; the bottom quark with negligible parton luminosity contributes very little. However, the branching fraction reduces compared to the flavor universal scenario and hence the total di-jet cross section decreases. 

Finally we note that in the narrow-width approximation the quantity $\xi$ can be determined by a measurement of the ratio of the cross sections where the resonance decays to top quark pairs or to jets:
\beq
\frac{ \sigma(pp\rightarrow C/\zp \rightarrow t\bar{t}) 	}{ \sigma(pp\rightarrow C/\zp \rightarrow jj)	} \simeq \frac{ Br(C/\zp\rightarrow t\bar{t})}{ Br(C/\zp\rightarrow jj)} = \frac{\xi^2}{ 4 + \xi^2}\,. 
\label{eq:xi}
\eeq
We introduce next a new variable to help distinguish color-octet and color-singlet states, described in this section, in the di-jet channel.  

\section{Color Discriminant Variable}
\label{sec:coldis}

In the previous section, we introduced generic color-octet and leptophobic color-singlet states that couple to SM quarks with differing strengths and color structures. A resonance of either type will be produced copiously at a hadron collider and will decay into two jets or top quark pairs. Decays in the top quark channel have the advantage of possible leptons in the final state and hence better reconstruction efficiencies compared to the SM QCD background. However they suffer from a smaller cross section (due to smaller branching fraction to top pairs), overall smaller efficiency due to the large number of final state particles and still a relatively large background. On the other hand the simple topology of the decay into two highly energetic central jets with a larger branching ratio and higher efficiency of jet selection makes the di-jet channel a discovery mode despite very large QCD backgrounds. Such resonances will be discovered in the di-jet channel as simple ``bumps" (in the narrow width approximation) over an exponentially falling QCD di-jet background. The channel with decays to top quark pairs, while suffering from a smaller rate, is nonetheless an important one as it helps determine flavor universal from flavor non-universal scenarios as described in Sec.~\ref{sec:resltnonuniv}. 

Once such a resonance has been discovered in the di-jet channel, it is possible to obtain a measurement of its cross section, mass and width. The question then arises about the nature of the resonance - is it a color-octet or a color-singlet resonance? An estimate of the couplings ($g_L^2 + g_R^2$) for each resonance can be obtained from the cross section measurement~\cite{Dobrescu:2013cmh}. This can possibly be used to eliminate some scenarios where the deduced couplings are either not motivated theoretically or excluded by other experiments. However, there will be large regions in parameter space where both color-octet and leptophobic color-singlet scenarios survive. To keep our study widely applicable we propose a model-independent approach to distinguish color-octet and color-singlet resonances that depends mainly on the kinematics of the process.

The cross section for the production and decay of a resonance in the narrow-width approximation can be written as 
\beq
\sigma(pp \stackrel{V} \longrightarrow jj) \simeq \sigma(pp \to V) Br(V \to jj),
\label{eq:genxsec}
\eeq
where $V$ is a generic resonance, $\sigma(pp \to V)$ is the cross section for producing the resonance and $Br(V \to jj)$ is the branching fraction for the resonance to decay to jets. Using this approximation and factoring out the couplings, color factors and mass dependence explicitly we can write the cross section for a color-octet as
\bea
\nn
\sigma(pp \to C \to jj) &=& \frac{4}{9}  \alpha_s \left( g_{C_L}^2 + g_{C_R}^2 \right) \frac{1}{M_C^2} \Sigma(pp \to C) Br(C \to jj)\\
&=& 	\frac{8}{9} \frac{\Gamma_C}{M_C^3} \Sigma(pp \to C) Br(C \to jj), 
\label{eq:colxsec}
\eea
where $\Sigma(pp \to C)$ is dependent only on the parton distribution functions (PDFs), kinematics, and phase space factors and the expression for the width of the flavor universal coloron in Eq.~(\ref{eq:colwiduniv}) has been used to obtain the final form of Eq.~(\ref{eq:colxsec}). Similarly, we can express the cross section for the leptophobic $\zp$ as
\bea
\nn
\sigma(pp \to \zp \to jj) &=& \frac{1}{3}  \alpha_w \left( g_{\zp_L}^2 + g_{\zp_R}^2 \right) \frac{1}{M_{\zp}^2}\Sigma(pp \to \zp) Br(\zp \to jj)\\
&=& 	\frac{1}{9} \frac{\Gamma_{\zp}}{M_{\zp}^3} \Sigma(pp \to \zp) Br(\zp \to jj), 
\label{eq:zpxsec}
\eea
where the expression for the width of the flavor universal leptophobic $\zp$ in Eq.~(\ref{eq:zpwiduniv}) has been used to obtain the final form of Eq.~(\ref{eq:zpxsec}). 

Suppose a new di-jet resonance is discovered with a particular cross section and mass; it is important to determine whether it is a coloron described by Eq.~(\ref{eq:colxsec}) or a $\zp$ described by Eq.~(\ref{eq:zpxsec}).  Comparing Eqs.~(\ref{eq:colxsec}) and (\ref{eq:zpxsec}) for equal $\sigma$ and equal $M$ and noting that $\Sigma$ and di-jet branching ratios are also equal in the two cases, we find the following relationship between the widths:
\beq
\Gamma^\ast_{\zp}
= {8 \Gamma^\ast_{C}\,} ,
\label{eq:widcompuniv}
\eeq
where the asterisk signifies that we are comparing bosons with equal production cross-sections ($\sigma$). This implies that if a resonance is discovered in the di-jet channel, a measurement of the width can point to the color-structure of the resonance discovered. 

We get a similar expression for the flavor non-universal case by replacing the expressions for the width and the branching fraction in Eqs.~(\ref{eq:colxsec}) and (\ref{eq:zpxsec}) by those in Eqs.~(\ref{eq:widnonuniv}) and (\ref{eq:brnonuniv}). For the flavor non-universal coloron we have
\bea
\nn
\sigma(pp \to C \to jj) &=& \frac{4}{9}  \alpha_s \left( g_{C_L}^2 + g_{C_R}^2 \right) \frac{1}{M_C^2} \Sigma(pp \to C) Br(C \to jj)\\
&=& 	\frac{8}{9} \frac{\Gamma_C}{M_C^3}\left(\frac{6}{4+2\xi^2}\right) \Sigma(pp \to C) \left(\frac{4 +\xi^2}{4+2\xi^2}\right), 
\label{eq:colxsecnon}
\eea
where $\Gamma_C$ is the width of the coloron in the flavor non-universal case. For the flavor non-universal leptophobic $\zp$ we have 
\bea
\nn
\sigma(pp \to \zp \to jj) &=& \frac{1}{3}  \alpha_w \left( g_{\zp_L}^2 + g_{\zp_R}^2 \right) \frac{1}{M_{\zp}^2}\Sigma(pp \to \zp) Br(\zp \to jj)\\
&=& 	\frac{1}{9} \frac{\Gamma_{\zp}}{M_{\zp}^3} \left(\frac{6}{4+2\xi^{\prime^2}}\right) \Sigma(pp \to \zp)  \left(\frac{4 +\xi^{\prime^2}}{4+2\xi^{\prime^2}}\right), 
\label{eq:zpxsecnon}
\eea
where $\Gamma_{\zp}$ is the width of the leptophobic $\zp$ in the flavor non-universal case. For a given $\xi$, which is determined by a measurement of the ratio of the cross sections as given in Eq.~(\ref{eq:xi}), we see that, for resonances with a mass $M$ and yielding equal total di-jet cross sections, the relation between the width of the coloron and the leptophobic $\zp$ in the flavor non-universal case remains the same as before:
\beq
\Gamma^\ast_{\zp}
= 8 \Gamma^\ast_{C} \,.
\label{eq:widcompnonuniv}
\eeq
We use this relation to introduce a new variable to distinguish color-octet and color-singlet resonances and parameterize it in a model-independent way. 

As discussed earlier, a discovery in the di-jet channel will inspire three immediate measurements - cross section, mass and width, which in turn depend on model dependent parameters such as couplings, color-structure and mass of the resonance. Note that the cross section is proportional to the color structure, square of the couplings and $M^{-2}$ while the width is proportional to square of the couplings and M. Denoting by $\sigma_{jj}$ the cross section for producing a resonance in the di-jet channel, we define
\beq
\dcol \equiv \frac{M^3}{\Gamma} \sigma_{jj} , 
\label{eq:dcol}
\eeq
as a color discriminant variable that is dimensionless by construction. This variable depends only on the color structure of the resonance being considered for a given $\xi$. For example, any two points in the parameter space of a coloron will lead to the {\it same} $\dcol$ for a given $\xi$ where as points in parameter space of a coloron and a leptophobic $\zp$ will lead to different values of $\dcol$ for the same $\xi$. Thus one can distinguish a color-octet and a color-singlet state in a relatively model-independent fashion {\it i.e.} without analyzing each point in parameter space separately. Next we discuss the constraints on the parameter space and the discovery potential of color-octet and color-singlet states at the LHC. 

\section{Parameter Space in di-jet Channel}
\label{sec:constr}

In this section we describe the region of parameter space in which using the color discriminant variable is applicable. First we discuss the current constraints on the parameter space of coloron and leptophobic $\zp$ models. Next we discuss the discovery prospect for colorons and leptophobic $\zp$s at the LHC with c.m. energy of 14 $\tev$, since the question of distinguishing the color structure of a resonance will arise only after the resonance has been discovered. The discovery and exclusion regions for the coloron and leptophobic $\zp$ are presented in Fig.~\ref{fig:allow}(a) and (b), respectively, for the flavor universal scenario and in Fig.~\ref{fig:allow}(c) and (d), respectively, for the flavor non-universal scneario.

\begin{figure}[t]
{
\includegraphics[width=0.495\textwidth, clip=true]{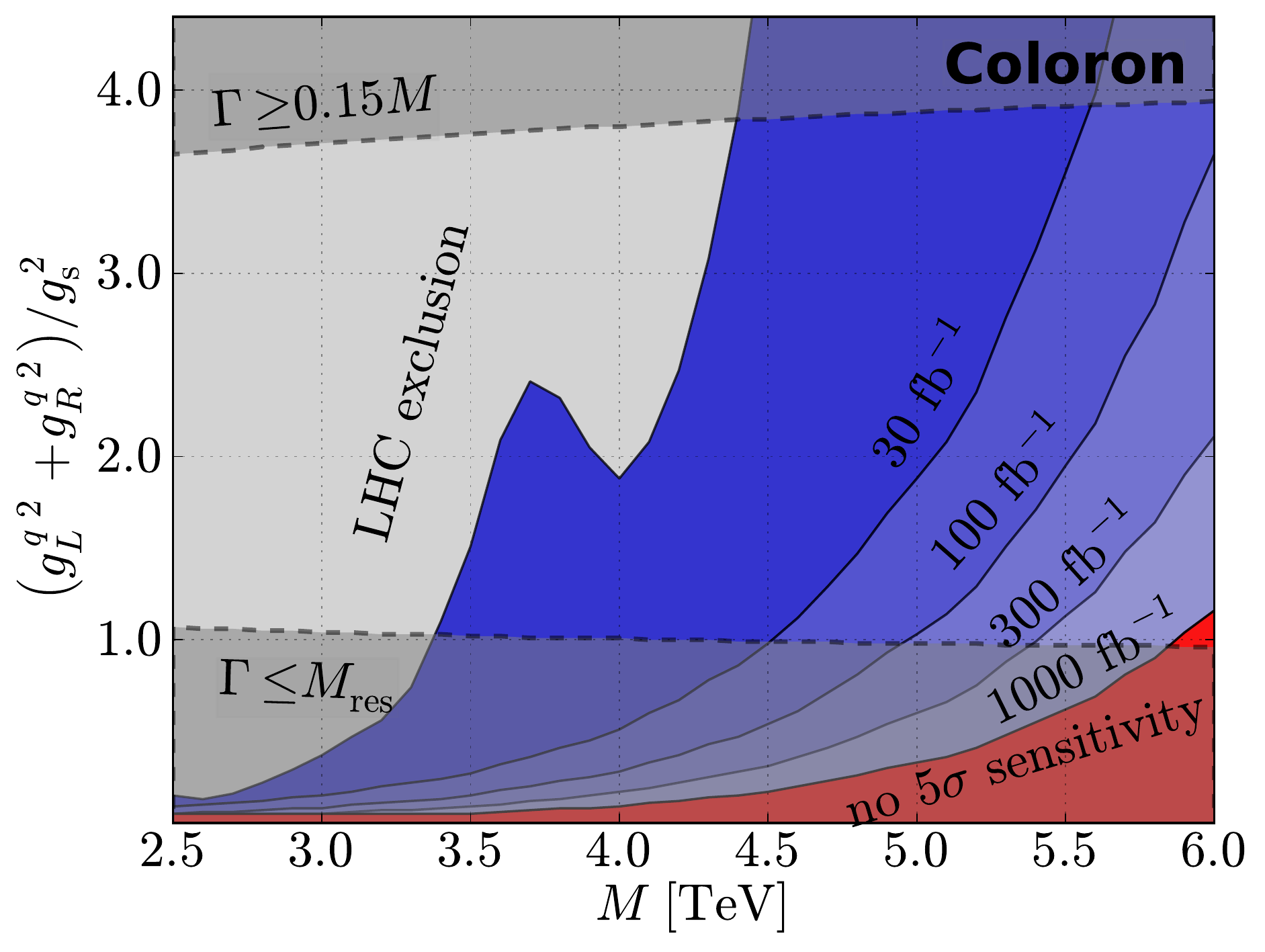}
\includegraphics[width=0.495\textwidth, clip=true]{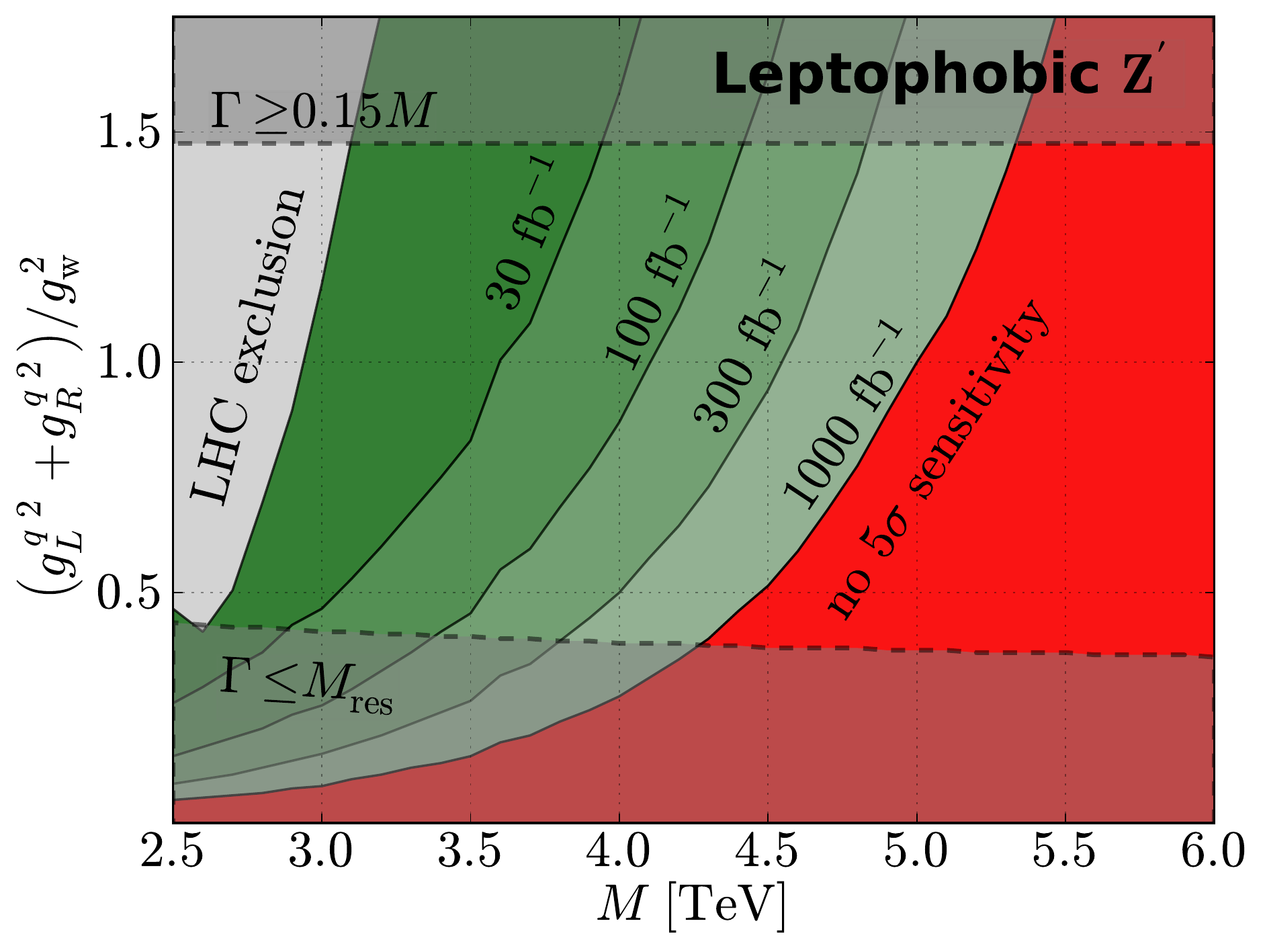}
\includegraphics[width=0.485\textwidth, clip=true]{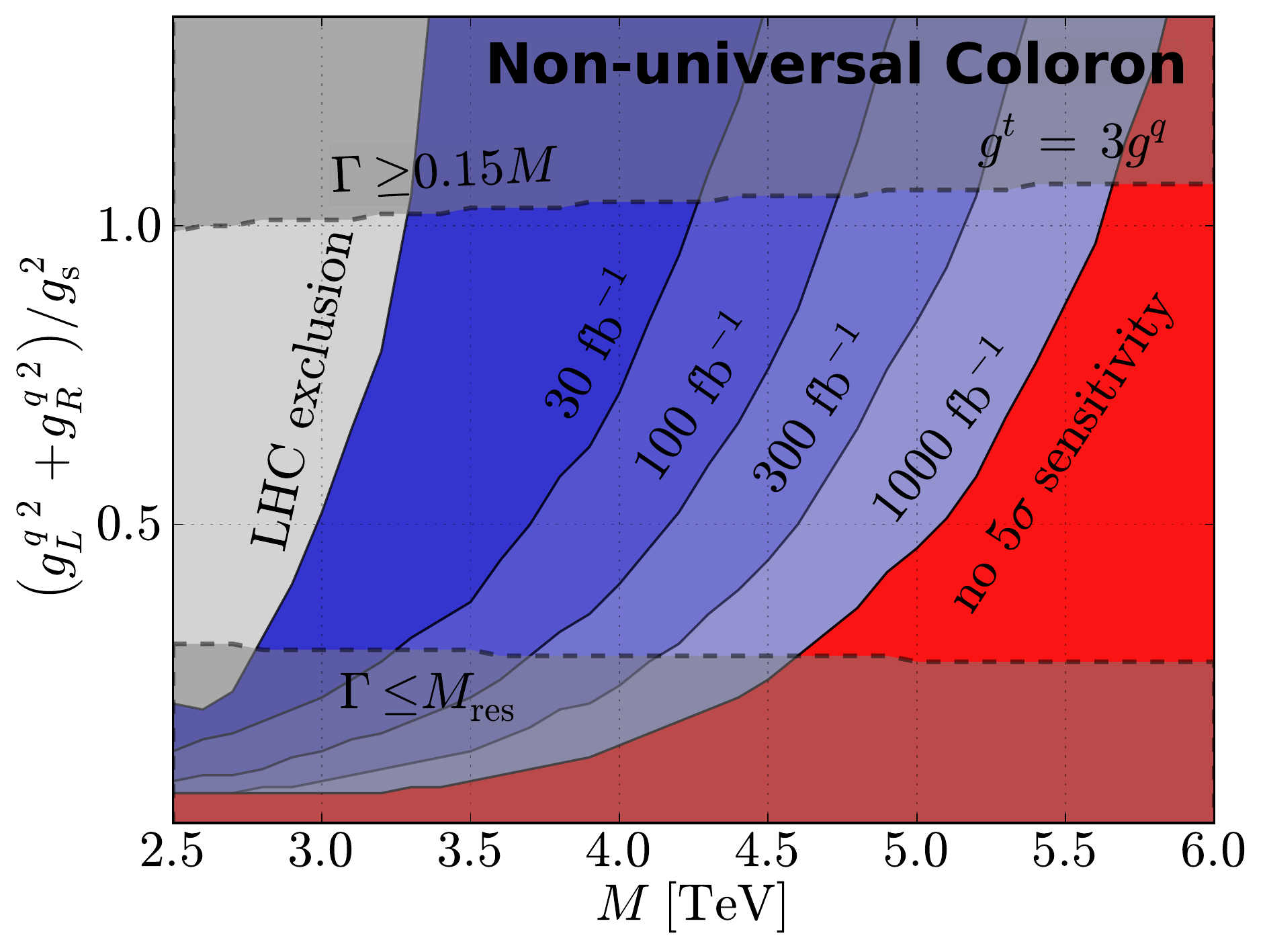}
\includegraphics[width=0.495\textwidth, clip=true]{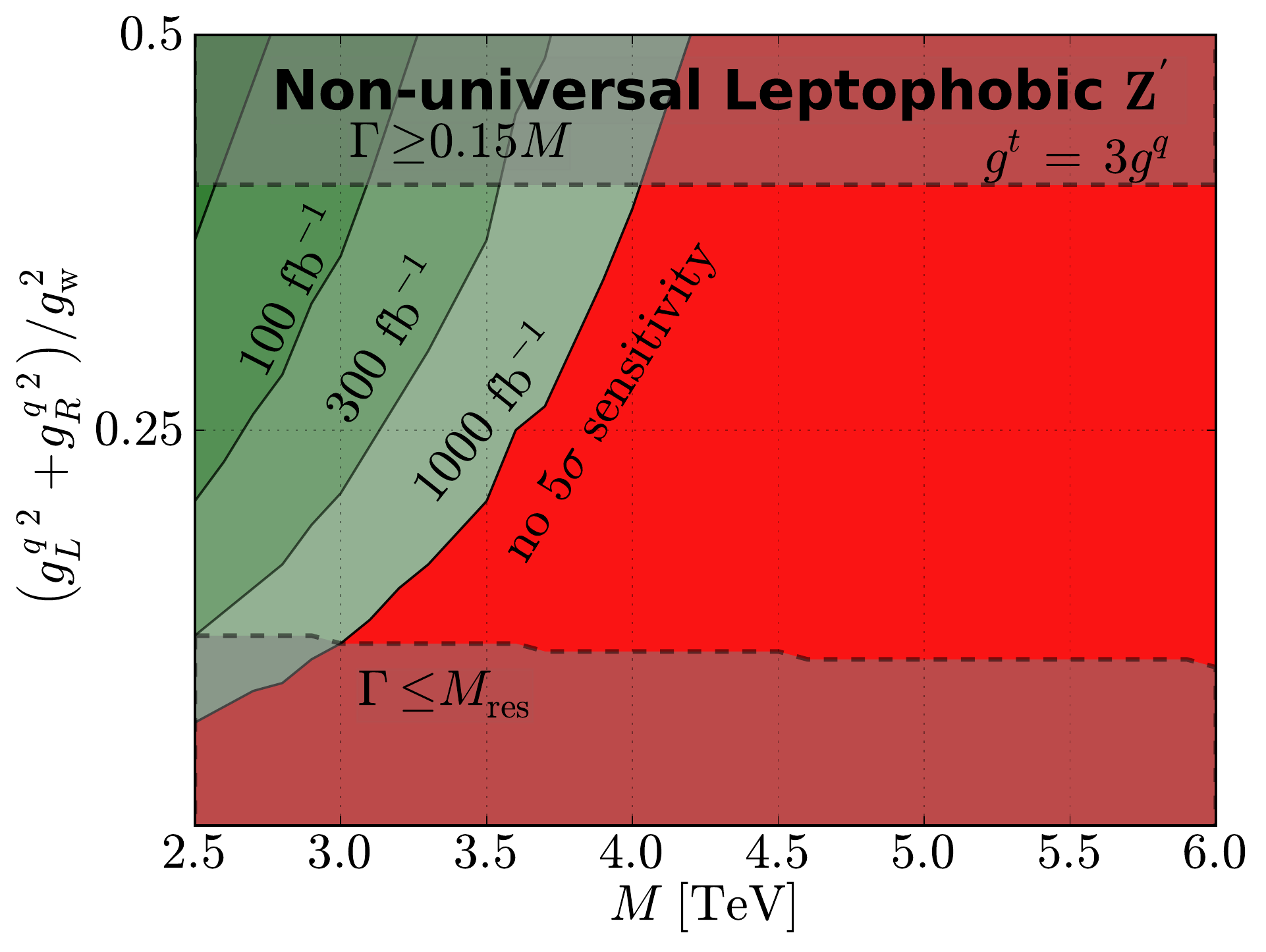}
}
\caption{(a) Top left: 5$\sigma$ discovery reach for a flavor universal coloron in the plane of the mass of the coloron (in $\tev$) and the square of the couplings at the LHC with $\sqrt{s} = 14\  \tev$. The discovery reach is shown in varying shades of blue for different luminosities ranging from 30 $\ifb$ to 1000 $\ifb$. The area marked ``no 5$\sigma$ sensitivity" corresponds to no discovery reach at 1000 $\ifb$ but may have some reach at higher luminosities. The area marked ``LHC exclusion" in gray corresponds to the exclusion from 8 $\tev$ LHC \cite{CMS:kxa}. The region above the dashed line marked $\Gamma \ge 0.15M$ corresponds to the region where the narrow-width approximation used in di-jet resonance searches is not valid \cite{Bai:2011ed, Haisch:2011up, Harris:2011bh}. The region below the horizontal dashed line marked $\Gamma \le \mres$ corresponds to the region where the experimental mass resolution is larger than the intrinsic width \cite{CMS:kxa}. See text for further details. (b) top right: same as (a) but for a leptophobic $\zp$ and the discovery reach is shown in varying shades of green. (c) bottom left: same as (a) but for the flavor non-universal coloron where $g_{C_{L,R}}^t = 3g_{C_{L,R}}^q.$ (d) bottom right: same as (b) but for the flavor non-universal $\zp$  with $g_{\zp_{L,R}}^t = 3g_{\zp_{L,R}}^q.$ 
}
\label{fig:allow}
\end{figure}

Searches for di-jet resonances where the width of the resonance is small compared to the mass have been carried out at the LHC by both CMS and ATLAS collaborations\footnote{
For a recent compilation of bounds on di-jet resonances, and their
interpretation in terms of resonance couplings, see Ref.~\cite{Dobrescu:2013cmh}. The results we present here, obtained from the experimental bounds cited, are consistent with the results of Ref.~\cite{Dobrescu:2013cmh}.}. They have found no evidence of such resonances and set exclusion limits on the product of cross section, branching ratio and acceptance for the 8 $\tev$ LHC run \cite{ATLAS:2012pu, Chatrchyan:2013qha, CMS:kxa}. They also present the theoretical estimate of the product of cross section times branching fraction for various sample models, including colorons and sequential $\zp$s. The acceptance for each model can then be estimated as a function of the mass without doing a full detector simulation. Using this estimated acceptance we translate the exclusion limits from Ref.~\cite{CMS:kxa} to obtain excluded regions in the plane of mass and coupling for the case of a coloron. The di-jet analysis \cite{CMS:kxa} presents results only for a sequential $\zp$ and hence we apply the acceptance for a sequential $\zp$ to the leptophobic $\zp$ as well. This is reasonable as the acceptance is dependent mainly on kinematics and not on the couplings and their structure at leading order. Note that in estimating the excluded regions we have used the most stringent results which come from CMS~\cite{CMS:kxa}. The results of this exercise are presented as gray regions labeled ``LHC exclusion" in Fig.~\ref{fig:allow} for the coloron and leptophobic $\zp$. Note that a larger region of parameter space is excluded for the coloron compared to the leptophobic $\zp$ due to the stronger coupling strength of the coloron to the SM quarks. On the other hand the excluded region for the flavor universal scenario is larger than that of the flavor non-universal scenario due to the smaller total cross section in the latter case as discussed earlier.

A resonance in the di-jet channel will be observed as a fluctuation in the exponentially falling QCD background. An estimation of the QCD di-jet background is notoriously difficult due to the reliance on leading order cross sections, uncertainties in the estimation of jet energy scale and other systematic uncertainties. Hence data driven methods are often employed to normalize the QCD background in a region away from the signal. In the absence of real data, a fit is performed to samples from full detector simulation to estimate the experimental sensitivity. As a full estimation of the QCD background and the sensitivity in the di-jet channel is beyond the scope (and focus) of this article, we use the results presented in Ref.~\cite{Gumus:2006mxa} to estimate the discovery potential of a coloron and a leptophobic $\zp$ in the di-jet channel at the LHC with $\sqrt{s} = 14\ \tev$. 

Similar to the case of current CMS studies, the authors of Ref.~\cite{Gumus:2006mxa} present the minimum cross section that can be observed at the LHC with $\sqrt{s} = 14\ \tev$ and for luminosities up to 10 $\ifb$. They present their discovery potential results as a product of cross section, branching ratio and acceptance for different masses of the resonance after taking into account statistical uncertainties (from background fluctuation) and systematic uncertainties. The systematic uncertainties include various sources such as jet energy scale, jet energy resolution, radiation and low mass resonance tails and luminosity. In addition they also present the theoretical estimate of the product of cross section times branching fraction for various sample models, including colorons and sequential $\zp$s. As before, the acceptance for each model can then be estimated as a function of mass and we translate the 5$\sigma$ discovery reach to regions in the plane of mass and coupling. The discovery reach from 10 $\ifb$ is then scaled appropriately to obtain the discovery reach for other values of  integrated luminosities, ${\cal L} = 30, 100, 300$ and $1000\  \ifb$. The regions that can be probed at the 5$\sigma$ level for the case of the flavor universal and flavor non-universal coloron are presented as regions of varying shades of blue in Fig.~\ref{fig:allow}(a) and Fig.~\ref{fig:allow}(c) respectively for the different luminosities listed above. For the leptophobic $\zp$, we use the acceptance for a sequential $\zp$ as explained earlier and the discovery reach is presented in Fig.~\ref{fig:allow}(b) and Fig.~\ref{fig:allow}(d) as different shades of green for the flavor universal and flavor non-universal case respectively. The region shown in red and labeled ``no 5$\sigma$ sensitivity" in Fig.~\ref{fig:allow}  corresponds to the case where the resonance will not be discovered at 5$\sigma$ with 1000 $\ifb$. Owing to the stronger coupling strength of the coloron to SM quarks the discovery reach for a coloron extends to larger masses while the reach for the leptophobic $\zp$ is limited to lower masses due to the weak coupling to SM quarks. Moreover, the discovery region for the flavor non-universal scenario shrinks compared to the flavor universal scenario due to the smaller total di-jet cross section in the former case as discussed earlier.

Note that the experimental search for resonances in the di-jet channel applies in the region where the narrow-width approximation is valid. The authors of Refs.~\cite{Bai:2011ed, Haisch:2011up, Harris:2011bh} point out that this approximation is valid only up to $\Gamma/M \le 0.15$. The area above the top dashed line in Fig.~\ref{fig:allow} indicates the region where the narrow-width approximation is not valid. Similarly, the measurement of the width of a resonance is limited by the experimental mass resolution $\mres$ and intrinsic widths smaller than $\mres$ cannot be distinguished. The area below the bottom dashed line in Fig.~\ref{fig:allow} indicates the region where the intrinsic width is smaller than the experimental resolution and cannot be distinguished. The estimate of the experimental mass resolution varies with mass and has been obtained from Ref.~\cite{CMS:kxa}. Hence the region where our analysis is applicable is between the two dashed lines and where there is discovery potential indicated by blue (green) colored regions for a coloron (leptophobic $\zp$). 

Finally we present some details about our simulation of signal samples. The production cross section was calculated using MadGraph5 \cite{Alwall:2011uj} and CTEQ6L1 PDFs \cite{Pumplin:2002vw} were used. The factorization and renormalization scales were set to be equal to the mass of the resonance. Next we discuss the sensitivity of the LHC to measure the color discriminant variable to distinguish color-octet and color-singlet states.

\section{Sensitivity at the LHC}
\label{sec:sens}

The color discriminant variable ($\dcol$) is a function of the mass ($M$) and intrinsic width ($\Gamma$) of the resonance as well as the cross section for producing the resonance in the di-jet channel ($\sigma_{jj}$). Hence statistical and systematic uncertainties in the measurement of the di-jet cross section, mass  and intrinsic width  of the resonance play a key role in the measurement of the color discriminant variable  and hence in distinguishing a color-octet from a color-singlet resonance. In this section we discuss the uncertainties in the measurement of the mass, intrinsic width and the cross section for producing the resonance in the di-jet channel and their effect on the uncertainties in the measurement of $\dcol$ at the LHC with $\sqrt{s} = 14\,\tev$. Motivated by current constraints and future prospects described in Sec.~\ref{sec:constr} we consider resonance masses in the range $2.5 - 6\ \tev$.

The uncertainty in the measurement of the di-jet cross section can be written as
\beq
\frac{\Delta\sigma_{jj} }{\sigma_{jj} }  = \frac{1}{\sqrt{N}} \oplus \varepsilon_{\sigma\,\rm{sys}}\,,
\eeq
where $N$ is the number of signal events, $\varepsilon_{\sigma\,\rm{sys}}$ is the fractional systematic uncertainty and $\oplus$ indicates that the uncertainties are added in quadrature. The discovery of a resonance in the di-jet channel is a pre-requisite for measuring $\dcol$. Hence in the rest of the article $N$ indicates the number of signal events required (above background fluctuation) to obtain a 5$\sigma$ discovery and has been obtained from Ref.~\cite{Gumus:2006mxa} as described in Sec.~\ref{sec:constr}. The sources of systematic uncertainties in measuring the di-jet cross section include 
 jet energy scale, jet energy resolution, radiation and low mass resonance tail and luminosity \cite{Ball:2007zza}. The effect of all these systematic uncertainties was estimated in Ref.~\cite{Gumus:2006mxa} and presented as a fractional uncertainty (as a function of the mass) normalized to the di-jet cross section required to obtain 5$\sigma$ discovery (above background fluctuation) at the LHC with $\sqrt{s} = 14\ \tev$. The fractional uncertainty ($\varepsilon_{\sigma \,\rm{sys}}$) varies from $0.28$ to $0.41$ in the mass range of interest and is listed in Table~\ref{tab:uncert}.

The uncertainty in the measurement of the di-jet mass is given by
\beq
\frac{\Delta M}{M} = \frac{1}{\sqrt{N}} \left[ \frac{\sigma_\Gamma}{M} \oplus \frac{\mres}{M} \right] \oplus \left(\frac{\Delta M}{M}\right)_{\rm{JES} } \,,
\eeq
where $\sigma_\Gamma$ is the standard deviation corresponding to the intrinsic width of the resonance ($\Gamma\simeq 2.35 \sigma_\Gamma$ assuming a Gaussian distribution), $\mres$ is the experimental di-jet mass resolution and $\left(\Delta M/M\right)_{\rm{JES}}$ is the uncertainty in the mass measurement due to uncertainty in the jet energy scale. The various components of systematic uncertainties contributing to the uncertainty in the mass measurement depend on each experiment and detector and their estimate for different experiments and c.m. energies are listed in Table~\ref{tab:uncert}. The specific values used in our analysis are indicated by an asterisk (*). 

The uncertainty in the measurement of the intrinsic width is given by
\beq
\frac{\Delta\Gamma}{\Gamma} =   \sqrt{ \frac{1}{2(N-1)}  \left[1 + \left(\frac{\mres}{\sigma_{\Gamma}}\right)^2 \right]^2
+  \left(\frac{\mres}{\sigma_\Gamma}\right)^4  \left( \frac{\Delta \mres}{\mres} \right)^2  } \,,
\label{eq:delta-gamma}
\eeq
where $\Delta \mres$ is uncertainty in the di-jet mass resolution due to uncertainty in the jet energy resolution. Again, the estimate of the various components contributing to the uncertainty in the width measurement are listed in Table~\ref{tab:uncert} for various experiments and the values used in the analysis are indicated by an asterisk(*). See Appendix~\ref{sec:app} for details on calculating the expression for uncertainty in the intrinsic width given in Eq.~(\ref{eq:delta-gamma}). 

The estimation of systematic uncertainties depends on detector details and the energy of the collider and an accurate estimate of any systematic uncertainty can be done only after the machine is operational and has been calibrated. However as the LHC is yet to run at $\sqrt{s} = 14\ \tev$, the energy for which we present our results, we have the choice of using the systematic uncertainties estimated using full detector simulation (but no real data) at $\sqrt{s} = 14\ \tev$ or of using the systematic uncertainties from real LHC data but for $\sqrt{s} = 8\ \tev$. We will use the estimate for systematic uncertainties from actual LHC data where available and assume that any future LHC run will be able to reach at least the current level of uncertainties, if not better. This is a reasonable assumption as experiments tend to make improvements in their estimation of errors and efficiencies with real data, compared to original estimates from simulated data, due to improved experimental techniques. For example, the uncertainty in mass due to the uncertainty in jet energy scale at the LHC with $\sqrt{s} = 14\, \tev$ was expected to be about $5\%$ \cite{Ball:2007zza} while the current analyses at $8\,\rm{TeV}$ show that an uncertainty of $1.25\%$ is achievable \cite{CMS:kxa}. Note that we have modeled all systematic uncertainties to be Gaussian and hence added them in quadrature. We have also not included any correlation between the uncertainties. Most of the systematic uncertainties were estimated for resonance masses up to 5 $\tev$ for the different experiments and we have extrapolated this estimate to resonance masses up to 6 $\tev$. To account for the possibility that the systematic uncertainties are larger than the ones we use in the analysis we also present our results for the case where all the systematic uncertainties used in evaluating $\dcol$ are increased by a factor of $1.5$. We believe this rather conservative estimate will be able to cover reasonable fluctuations in systematic uncertainties due to higher energy, larger luminosity and other effects not included in our study. 

\begin{table}[!h]
\begin{tabular}{| c | c | c |  c  | c | r |}
\hline
\hline
	Systematic Uncertainty		&
	Notation		&
	Value		&
	Mass Range	&
	$\sqrt{s}$	&
	Experiment\\
\hline
	Di-jet cross section &
	\multirow{2}{*}{ \large{$\varepsilon_{\sigma \,\rm{sys}}$} }	&
	\multirow{2}{*}{$0.28-0.41^\ast$} &
	\multirow{2}{*}{$2.5-6\,\tev$}	&
	\multirow{2}{*}{$14\,\tev$}		&
	\multirow{2}{*}{LHC \cite{Gumus:2006mxa} \hspace{-0.14cm}}\\ 
	 uncertainty (fractional) &
		&
	 	&
	 	&
		&
	\\
\hline
	\multirow{3}{*}{Mass resolution}	&
	\multirow{3}{*}{\large {$\frac{\mres}{M}$}	}&
	$0.045 - 0.035^\ast$ &
	 $2.5-6\,\tev$	&
	$8\,\tev$		&
	CMS \cite{CMS:kxa}
\\
	{}	&
	{}	&
	$0.045  - 0.031$ &
	 $2.5-6\,\tev$	&
	$8\,\tev$		&
	ATLAS \cite{ATLAS:2012qjz}
\\
		&
	{}	&
	$0.071 - 0.062$ &
	 $2.5-6\,\tev$	&
	$14\,\tev$		&
	LHC \cite{Gumus:2006mxa}
\\
\hline
	Mass resolution 	&
	\multirow{2}{*}{ \large {$\frac{\Delta \mres} {M}$}  }	&
	$0.1^\ast$ &
	any			&
	$8\,\tev$		&
	CMS \cite{Chatrchyan:2013qha}
\\
	uncertainty	&
		&
	$0.1$ &
	any			&
	$14\,\tev$		&
	LHC \cite{Gumus:2006mxa}
\\
\hline
	Mass uncertainty 	&
	\multirow{3}{*}{$\left( \frac{\Delta M}{M}\right)_{\rm{JES}}$} 	&
	$0.013^\ast$ &
	any	&
	$8\,\tev$		&
	CMS \cite{Chatrchyan:2013qha}
\\
	from jet energy	&
		&
	$0.028$ &
	any	&
	$8\,\tev$		&
	ATLAS \cite{ATLAS:2012qjz}
\\
	scale (JES)	&
		&
	$ 0.035$ &
	any	&
	$14\,\tev$		&
	LHC \cite{Gumus:2006mxa}
\\
\hline
\hline
\end{tabular}
\caption{Sources of systematic uncertainty contributing to uncertainties in measurement of the cross section, mass and width of a resonance in the di-jet channel at various experiments and c.m. energies. The values used in this analysis are indicated by an asterisk.
}
\label{tab:uncert}
\end{table}

\section{Results}
\label{sec:reslt}

In this section we describe the sensitivity of the LHC to distinguish a color-octet resonance from a color-singlet resonance using the color discriminant variable introduced in Sec.~\ref{sec:coldis}. We estimate the sensitivity at the LHC by evaluating $\dcol$ and include the uncertainties in estimating $\dcol$ as described in Sec.~\ref{sec:sens}. Motivated by current constraints on di-jet resonances and future sensitivity of the LHC to discover di-jet resonances as described in Sec.~\ref{sec:constr}, we consider the mass range of $2.5 - 6\ \tev$. We will present our results for the flavor universal and the flavor non-universal case at the LHC with $\sqrt{s} = 14\ \tev$ for varying integrated luminosities, namely, ${\cal L} = 30, 100, 300\ \rm{and}\ 1000\ \ifb$. 

\subsection{Flavor Universal Scenario}
\label{sec:resltuniv}

The sensitivity of the LHC with c.m. energy of $14\ \tev$ to distinguish color-octet and color-singlet resonances in the flavor universal scenario is shown in Fig.~\ref{fig:sensuniv}(a) - (d) for varying luminosities, ${\cal L} = 30, 100, 300\ \rm{and}\ 1000\ \ifb$. The sensitivity is presented in the plane of the mass of the resonance ($M$) in $\tev$ and the log of the color discriminant variable $(\dcol)$. In each panel, there are two separate bands, corresponding to a coloron and a leptophobic $\zp$. For each resonance, the uncertainty in the measurement of $\dcol$ due to uncertainties in the measurement of the cross section, mass and width of the resonance is indicated by gray bands around the central value of $\dcol$ represented as a black dashed line. The outer (dark gray) band corresponds to the uncertainty  in $\dcol$ when the width is equal to the mass resolution, {\it i.e.} $\Gamma = \mres$. A determination of the intrinsic width is not possible when $\Gamma < \mres$. The inner (light gray) band corresponds to the case where the width $\Gamma = 0.15M$. The narrow width approximation used in di-jet searches is not valid when $\Gamma > 0.15M$. Resonances with width $\mres \le \Gamma \le 0.15M$ will have bands that extend between the outer and inner gray bands. The blue (green) colored region indicates the region in parameter space of the coloron (leptophobic $\zp$) that has not been excluded by current searches~\cite{CMS:kxa} and has the potential to be discovered at the LHC at a $5\sigma$ level as described in Sec.~\ref{sec:constr}.  

\begin{figure}[t]
{\includegraphics[width=0.495\textwidth, clip=true]{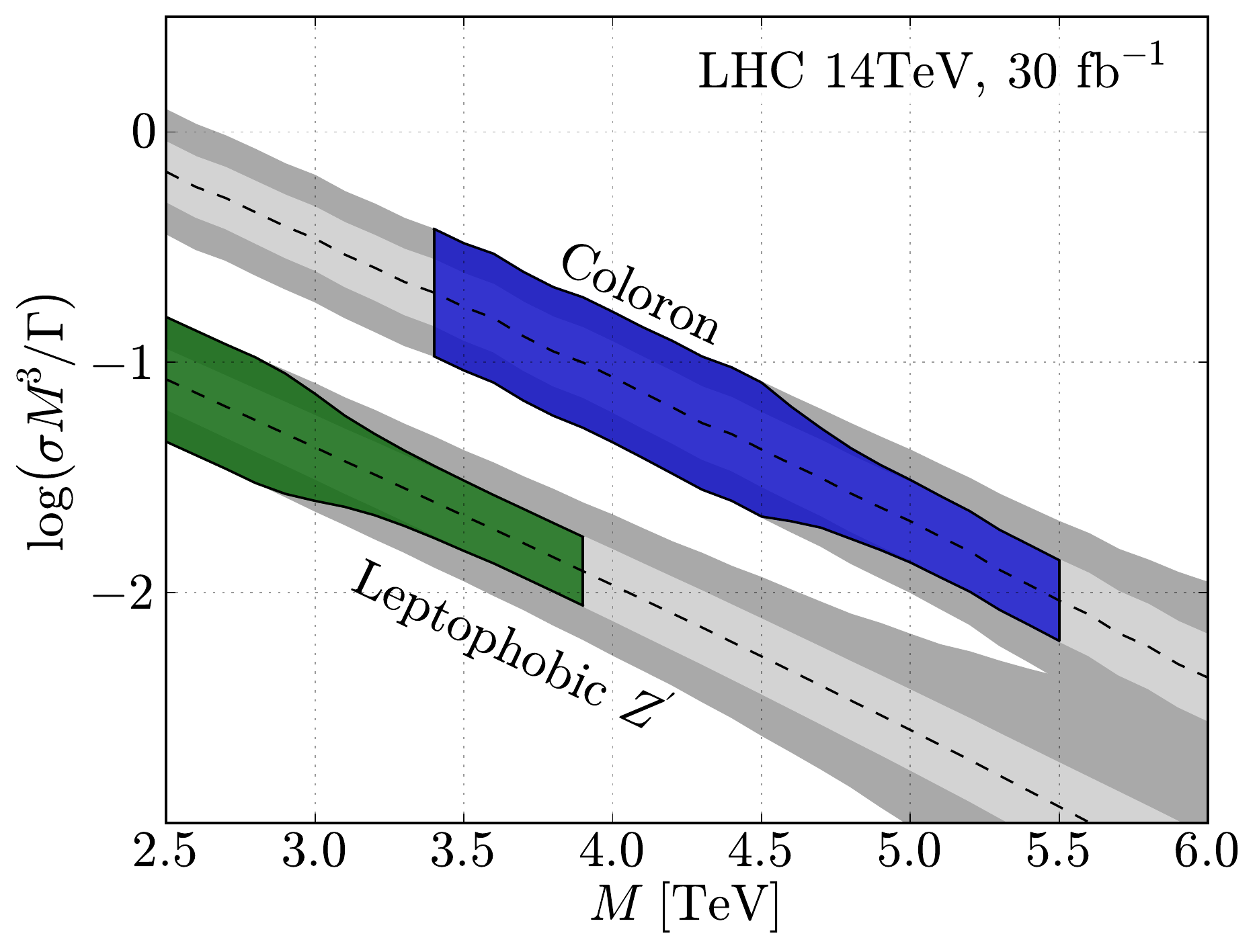}
\includegraphics[width=0.495\textwidth, clip=true]{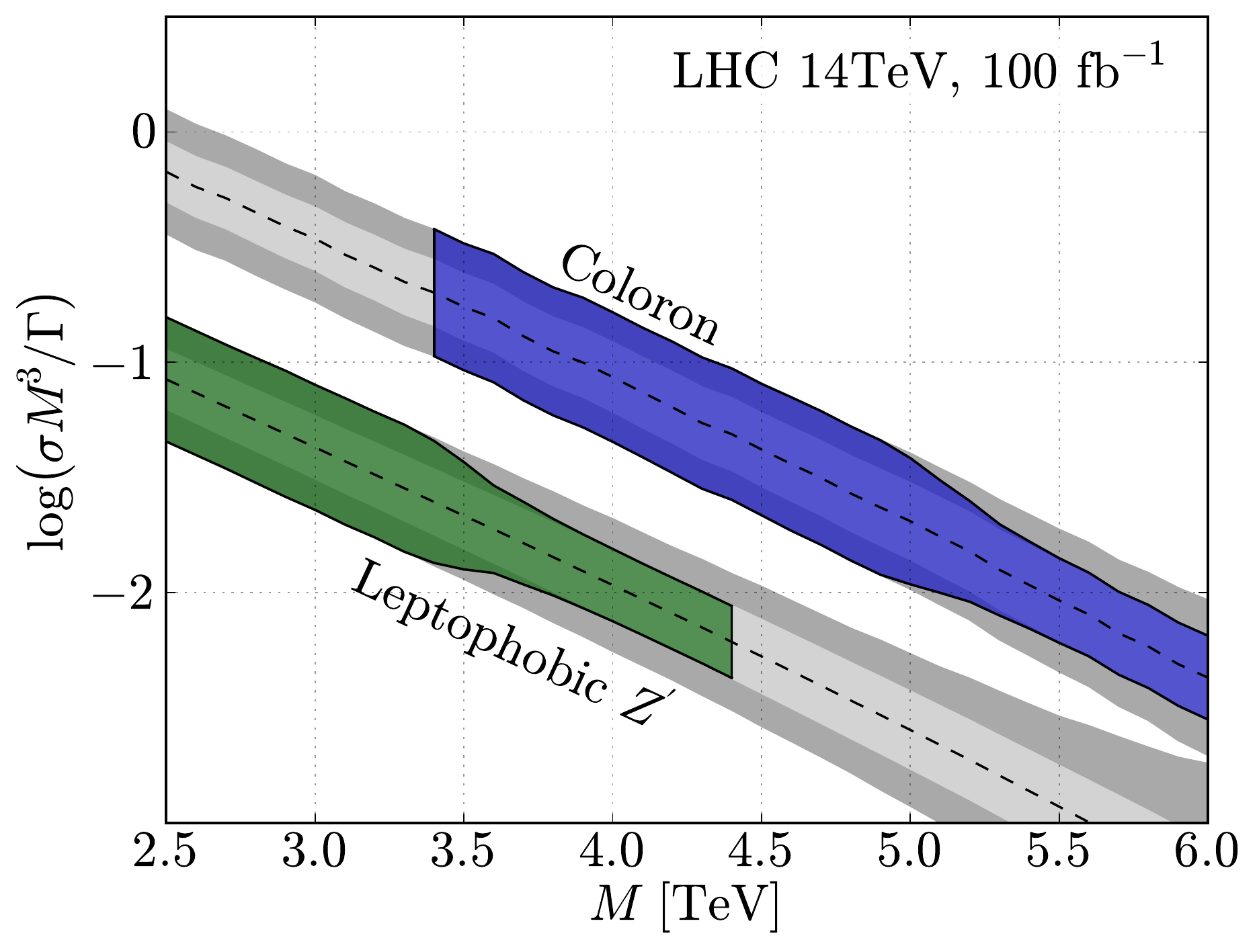}
\includegraphics[width=0.495\textwidth, clip=true]{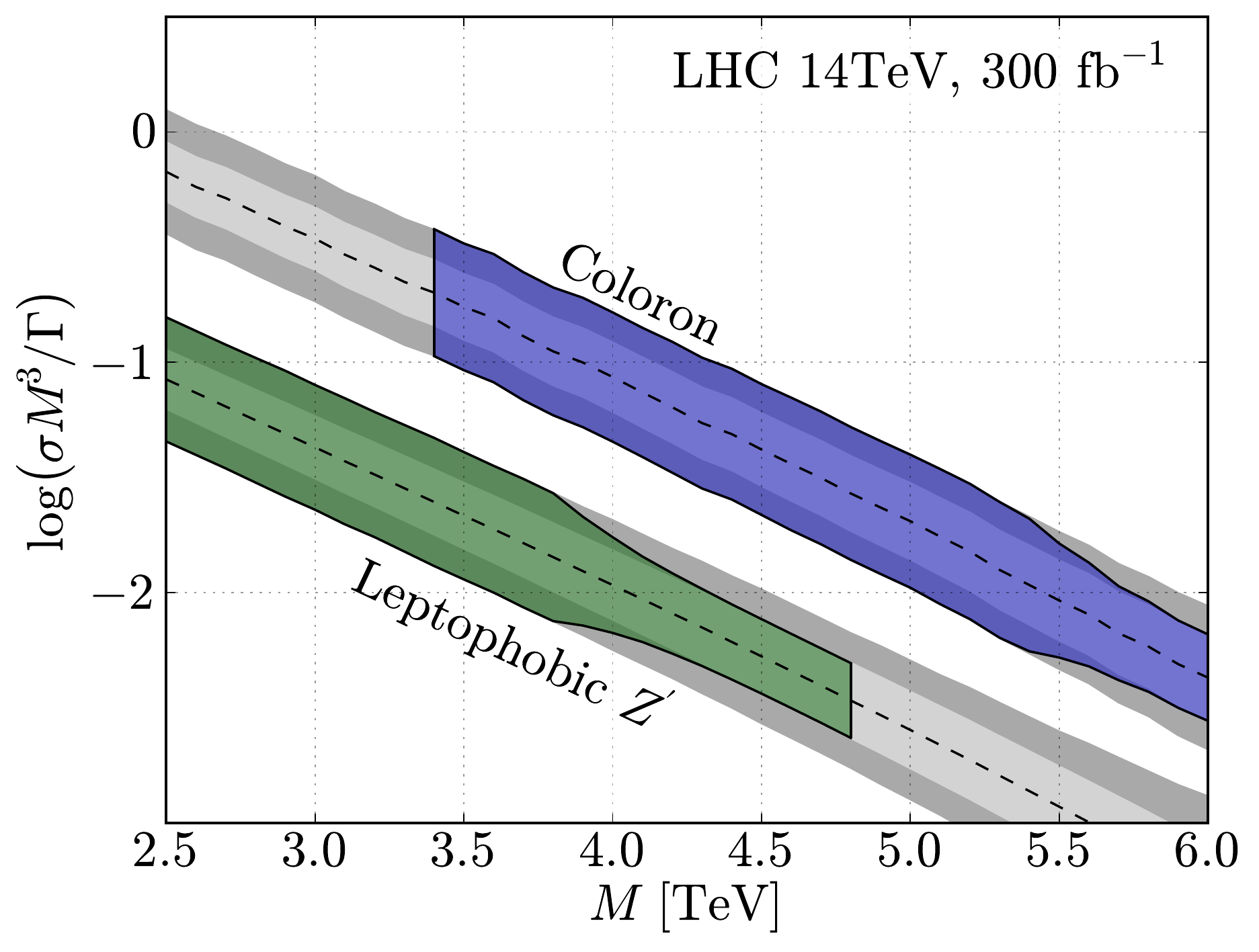}
\includegraphics[width=0.495\textwidth, clip=true]{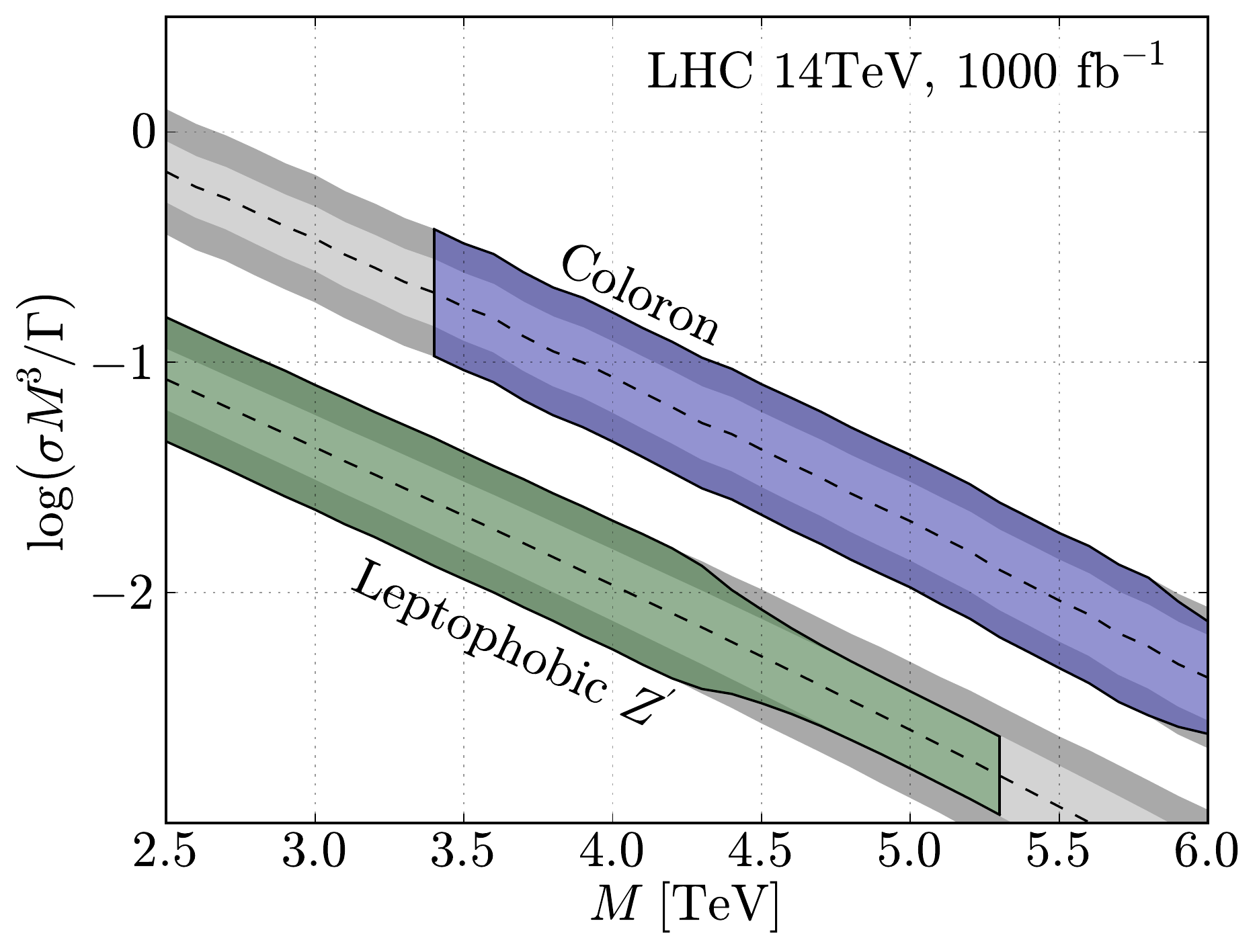}}
\caption{ (a) Top left: Sensitivity at the LHC with $\sqrt{s}=14\ \tev$ and integrated luminosity of $30\,\ifb$ for distinguishing a coloron from a leptophobic $Z^\prime$ in the plane of the log of the color discriminant variable $\left(\dcol =  \frac{M^3}{\Gamma} \sigma_{jj}\right)$ and mass (in TeV) for the flavor universal scneario. The central value of $\dcol$ for each particle is shown as a black dashed line. The uncertainty in the measurement of $\dcol$ due to the uncertainties in the measurement of the cross section, mass and width of the resonance is indicated by gray bands. The outer (darker gray) band corresponds to the uncertainty  in $\dcol$ when the width is equal to the experimental mass resolution {\it i.e.} $\Gamma = \mres$. The inner (lighter gray) band corresponds to the case where the width $\Gamma = 0.15M$. Resonances with width $\mres \le \Gamma \le 0.15M$ will have bands that extend between the outer and inner gray bands. The blue (green) colored region indicates the region in parameter space of the coloron (leptophobic $\zp$) that has not been excluded by current searches~\cite{CMS:kxa} and has the potential to be discovered at a $5\sigma$ level at the LHC with $\sqrt{s} = 14\ \tev$ after statistical and systematic uncertainties are taken in to account. (b) Top right: Same as (a) but for an integrated luminosity of $100\ \ifb$. (c) Bottom left:  Same as (a) but for an integrated luminosity of $300\ \ifb$ (d) Bottom right: Same as (a) but for an integrated luminosity of $1000\ \ifb$. Note that the colored regions in all panels correspond to the same colored regions in the mass and coupling plane used in Fig.~\ref{fig:allow} for different luminosities.}
\label{fig:sensuniv}
\end{figure}

The results in Fig.~\ref{fig:sensuniv} illustrate several features. First, note that the bands for coloron and leptophobic $\zp$ are well separated vertically. This implies that the color discriminant variable is able to clearly distinguish between a color-octet and a color-singlet at the LHC after all uncertainties have been taken into account. The mass range (from $2.5 - 6\ \tev$) can be roughly divided into three regions: the low mass region where the $\zp$ band is green but the coloron band is grayed out; the high mass region where at least one band is grayed out, and an intermediate region where the coloron band is blue and the $\zp$ band is green. The color discriminant variable can be used to distinguish colorons and leptophobic $\zp$s in the intermediate region where both resonances are allowed, discoverable at the LHC, and have widths in the appropriate range ($\mres \le \Gamma \le 0.15M$). Note that this overlap region expands with increasing luminosity.  

Our analysis is not useful in the low mass region because a coloron there is either already excluded by LHC searches or is too narrow.  For example, consider a future discovery of a resonance in the di-jet channel at a mass of $3.0\ \tev$ and $100\ \ifb$ of integrated luminosity.  This could certainly correspond to a $\zp$ lying within the green band of Fig.~\ref{fig:sensuniv}(b).  However, from the lower left corner of Fig.~\ref{fig:allow}(a) we also see that it could correspond to a very narrow coloron, with a width less than the detector resolution. Being unable to measure the width of such a coloron accurately, the uncertainty in the measurement of our discriminant variable $\dcol$ would be very large and hence we would not be able to distinguish between a coloron and a leptophobic $\zp$.  

Now contrast this with the high mass region where the coloron has the potential to be discovered while the leptophobic $\zp$ band is grayed out. In this case if a resonance is discovered in the di-jet channel with a mass of $5.0\ \tev$ with $100\ \ifb$ integrated luminosity, then we would have confidence that it is a coloron and not a leptophobic $\zp$. A discovery in the di-jet channel (with the current analyses) requires the width to be relatively narrow ($\Gamma \le 0.15M$), while the width of a corresponding $\zp$ would be very broad as seen in Fig.~\ref{fig:allow}(b). Note that there are no current experimental strategies to discover very broad resonances and it is also beyond the scope of our analysis.

The color discriminant variable is a dimensionless quantity that depends only on the color factors for a given $\xi$. All points in the parameter space of a coloron and leptophobic $\zp$ give the same value of $\dcol^C$ and $\dcol^{\zp}$ respectively and $\dcol^C \ne \dcol^{\zp} $ for a given $\xi$. However the results in Fig.~\ref{fig:sensuniv} show a clear dependence of $\dcol$ on the mass of the resonance ($M$). This dependence on the mass is an artifact of the implicit dependence on mass coming from PDFs. A resonance with a large (small) mass corresponds to parton luminosity at large (small) $x$  and hence small (large) cross section. Note that at a fixed collision energy (at the parton level) the value of $\dcol$ for a given resonance will be universal and have no mass dependence. 

The uncertainties in the estimation of $\dcol$ are large for higher mass resonances compared to the ones at low mass. This is easily understood as high mass resonances have smaller cross sections and hence fewer signal events leading to large uncertainties. As the luminosity increases from $30\ \ifb$ to $1000\ \ifb$ the uncertainty at higher masses decreases due to the larger number of signal events and the width of the uncertainty bands becomes uniform. This is because systematic uncertainty will become progressively dominant when the number of events is sufficiently large that a $5\sigma$ discovery is possible.

\begin{figure}[t]
{\includegraphics[width=0.495\textwidth, clip=true]{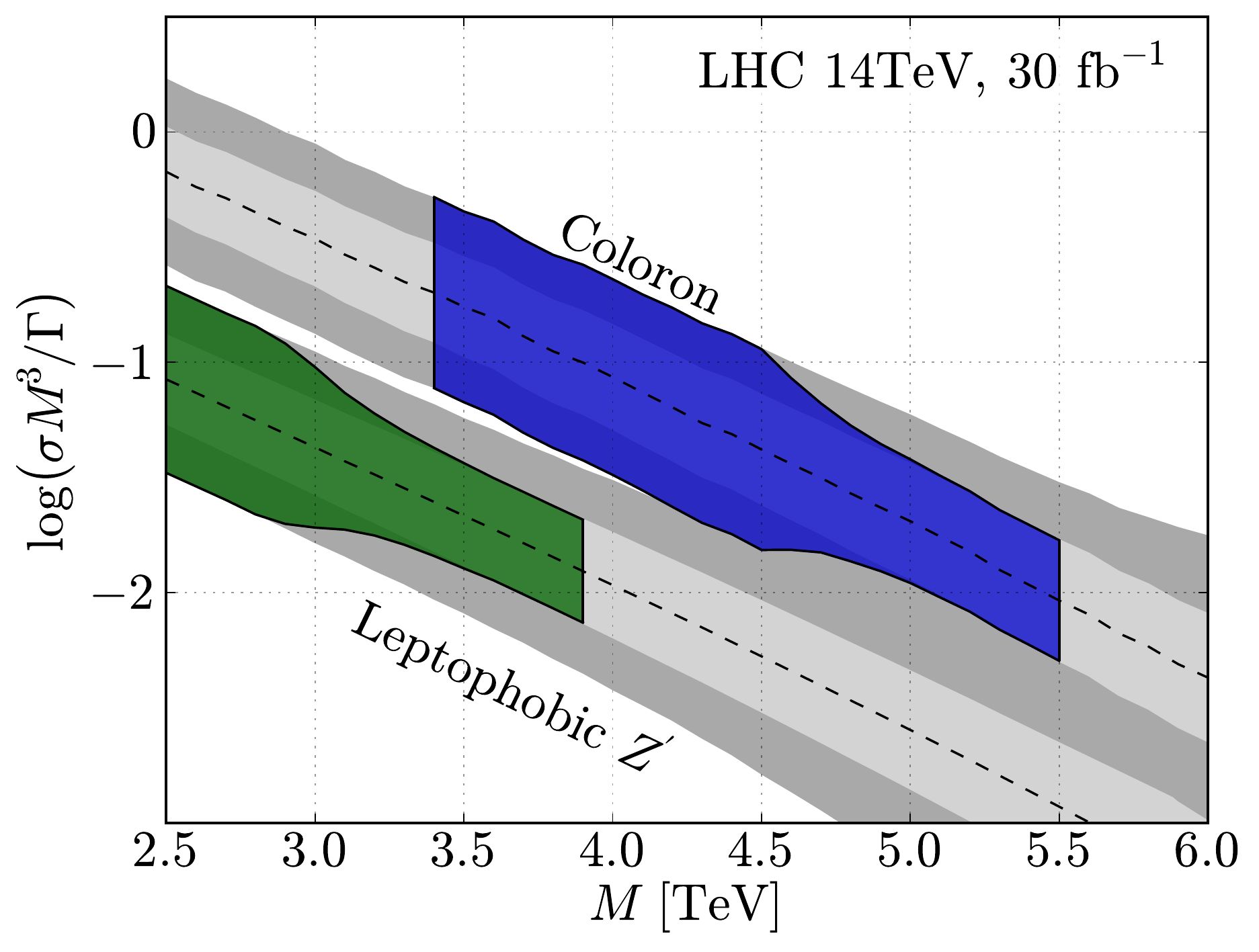}
\includegraphics[width=0.495\textwidth, clip=true]{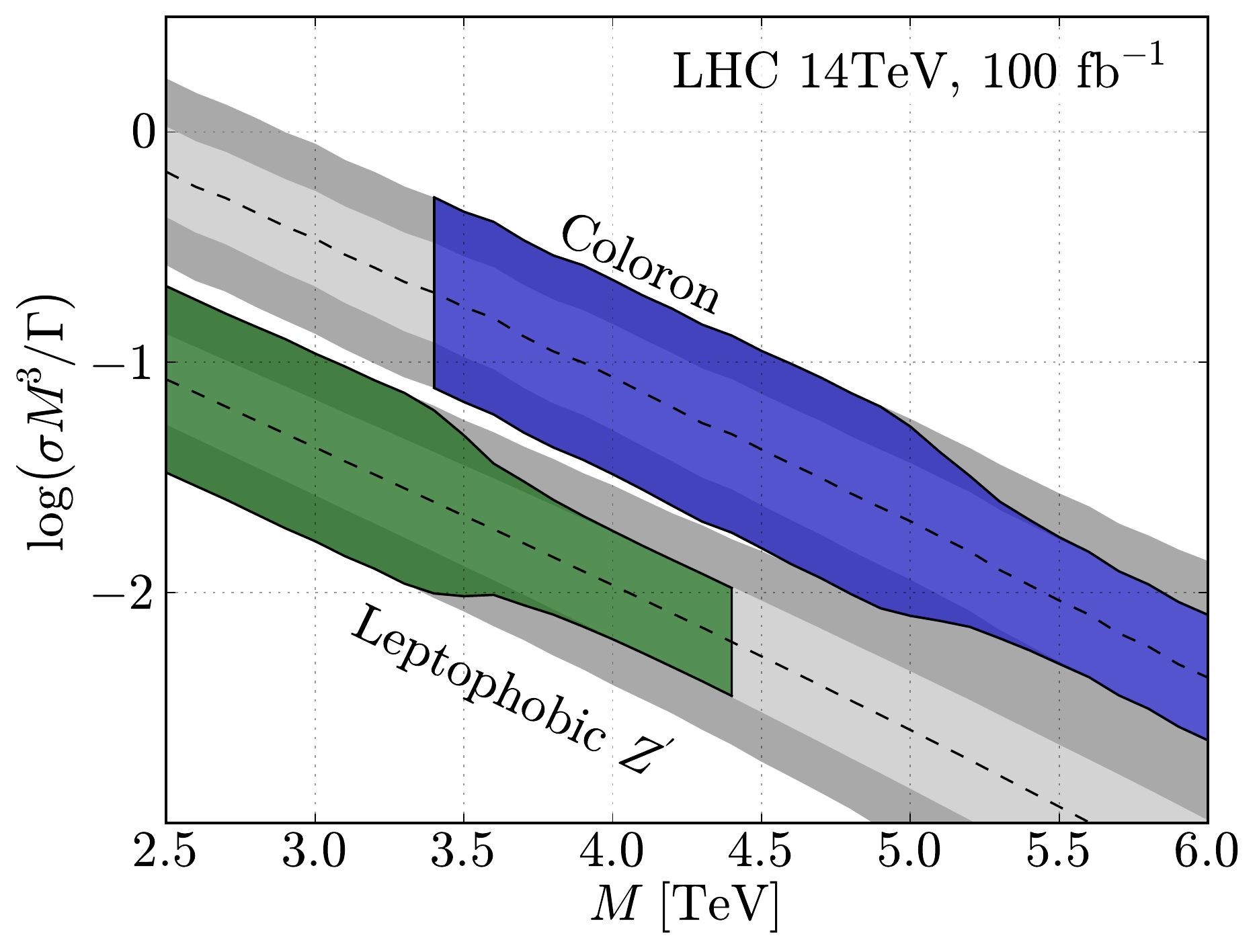}
\includegraphics[width=0.495\textwidth, clip=true]{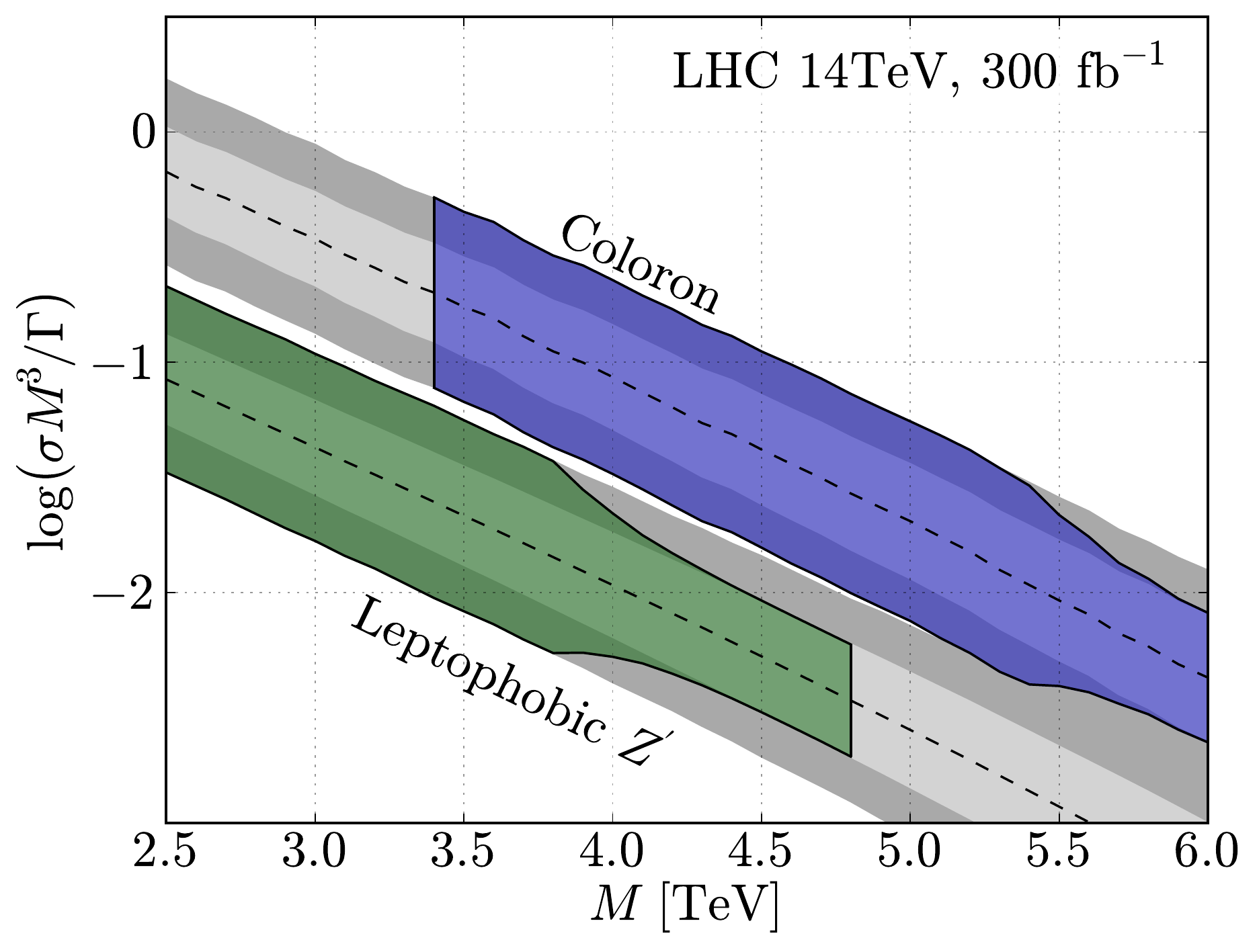}
\includegraphics[width=0.495\textwidth, clip=true]{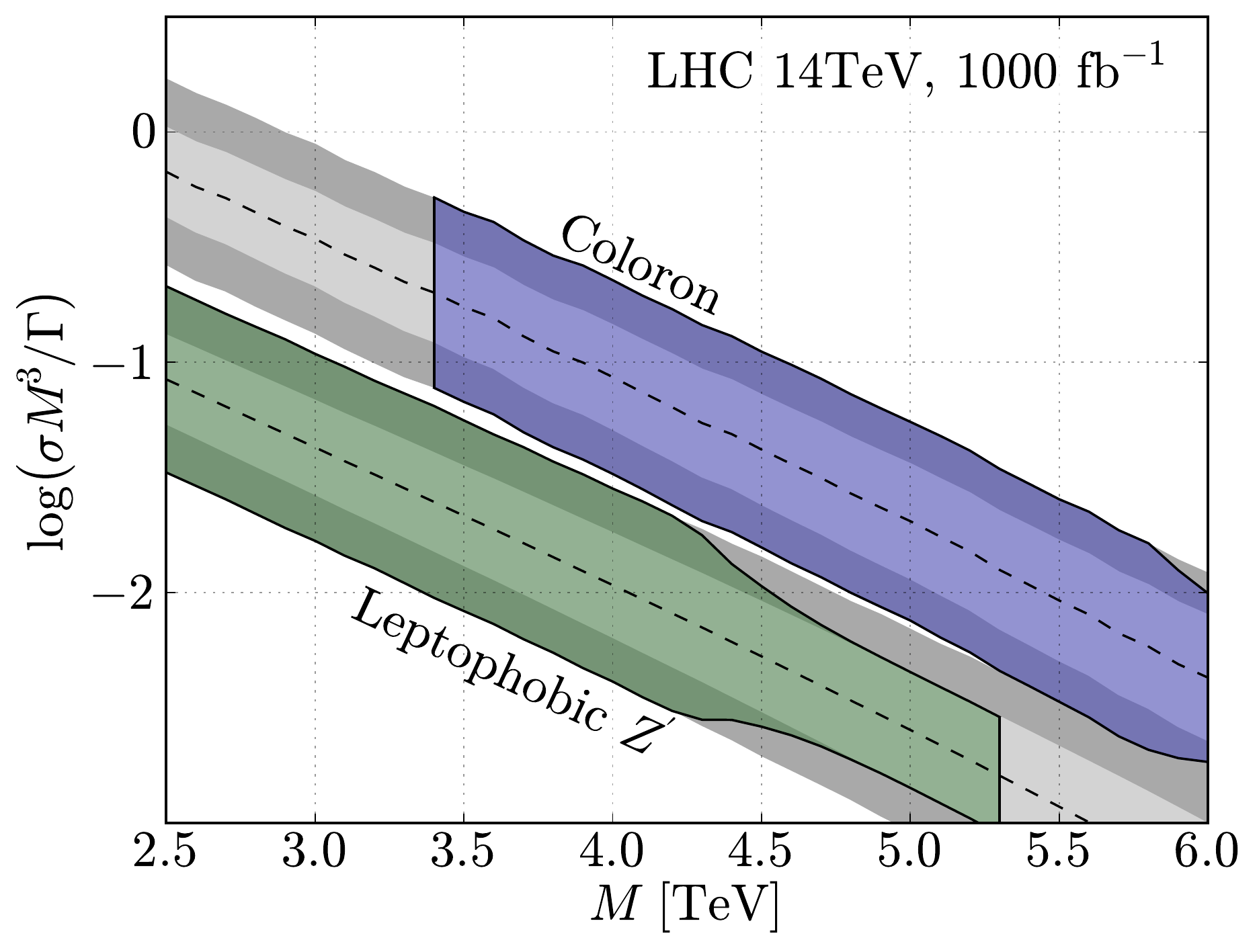}}
\caption{Same as Fig.~\ref{fig:sensuniv}, but with the systematic uncertainties in $M$, $\Gamma$ and $\sigma_{jj}$ increased by a factor of 1.5.}
\label{fig:sensuniv-xtra}
\end{figure}

The uncertainty in the color discriminant variable is smaller for the case of larger width (inner gray band) and is larger for the case of smaller width (outer gray band) resonances. This is because smaller width corresponds to smaller couplings and hence smaller cross section which leads to larger statistical uncertainty. In addition the uncertainty in the intrinsic width is inversely proportional to $\Gamma/\mres$ and hence leads to larger systematic uncertainty for smaller widths. This relation between the uncertainty in the width and the mass resolution is described in Appendix~\ref{sec:app}.  In addition, the size of the colored band (which corresponds to the discovery reach) is smaller for larger masses. This is because the couplings corresponding to a width of $\Gamma = \mres$ (outer region of the colored band) are small leading to a smaller cross section and hence not enough signal events for a $5\sigma$-discovery at a given mass. The larger couplings which correspond to larger widths (inner region of the colored band) and larger cross sections have the potential for a $5\sigma$ discovery. 

In figure \ref{fig:sensuniv-xtra}, we display the results for the scenario where systematic uncertainties in the measurement of mass, width and cross section of the resonance in the di-jet channel listed in Sec.~\ref{sec:sens} were increased by a factor of $1.5$. Note that our results remain robust even with this very conservative estimate of systematic uncertainties and the color-octet and color-singlet states can still be distinguished from each other.

\subsection{Flavor Non-universal Scenario}
\label{sec:resltnonuniv}

In this section we discuss the sensitivity of the LHC with c.m. energy of $14\ \tev$ to distinguish between color-octet and color-singlet resonances in the flavor non-universal scenario. As illustrative examples we present the results for two different flavor non-universal scenarios in Fig.~\ref{fig:sensnon} for the mass range $2.5 - 6\ \tev$ and for various couplings and widths. In Fig.~\ref{fig:sensnon}(a) and Fig.~\ref{fig:sensnon}(b) we present the results in the plane of mass of the resonance and the log of the color discriminant variable for $g_{L,R}^t = 2g_{L,R}^q$ for integrated luminosities of $300\ \ifb$ and $1000\ \ifb$ respectively. In Fig.~\ref{fig:sensnon}(c) and Fig.~\ref{fig:sensnon}(d) we present similar results for $g_{L,R}^t = 3g_{L,R}^q$ for integrated luminosities of $300\ \ifb$ and $1000\ \ifb$ respectively.

\begin{figure}[t]
{\includegraphics[width=0.495\textwidth, clip=true]{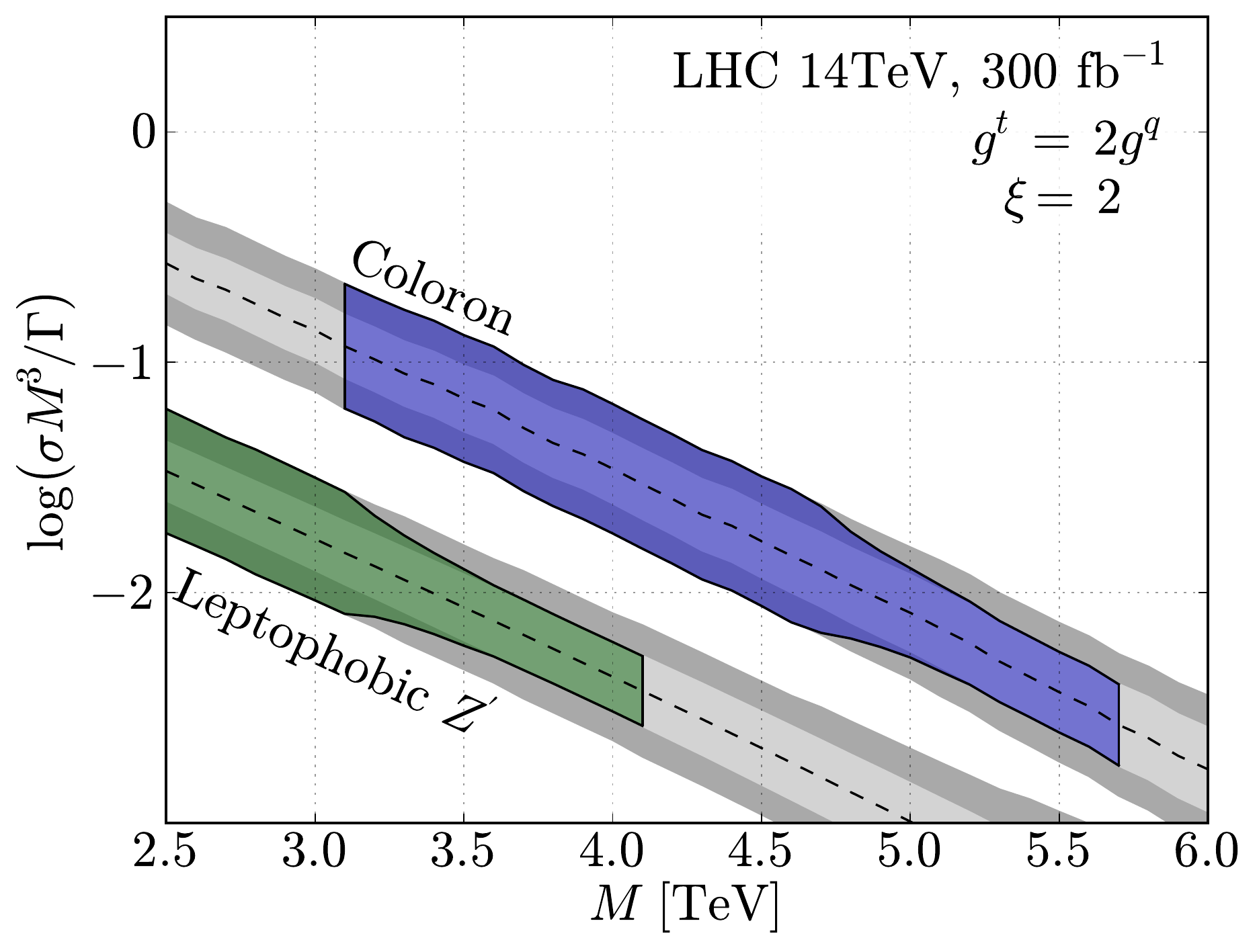}
\includegraphics[width=0.495\textwidth, clip=true]{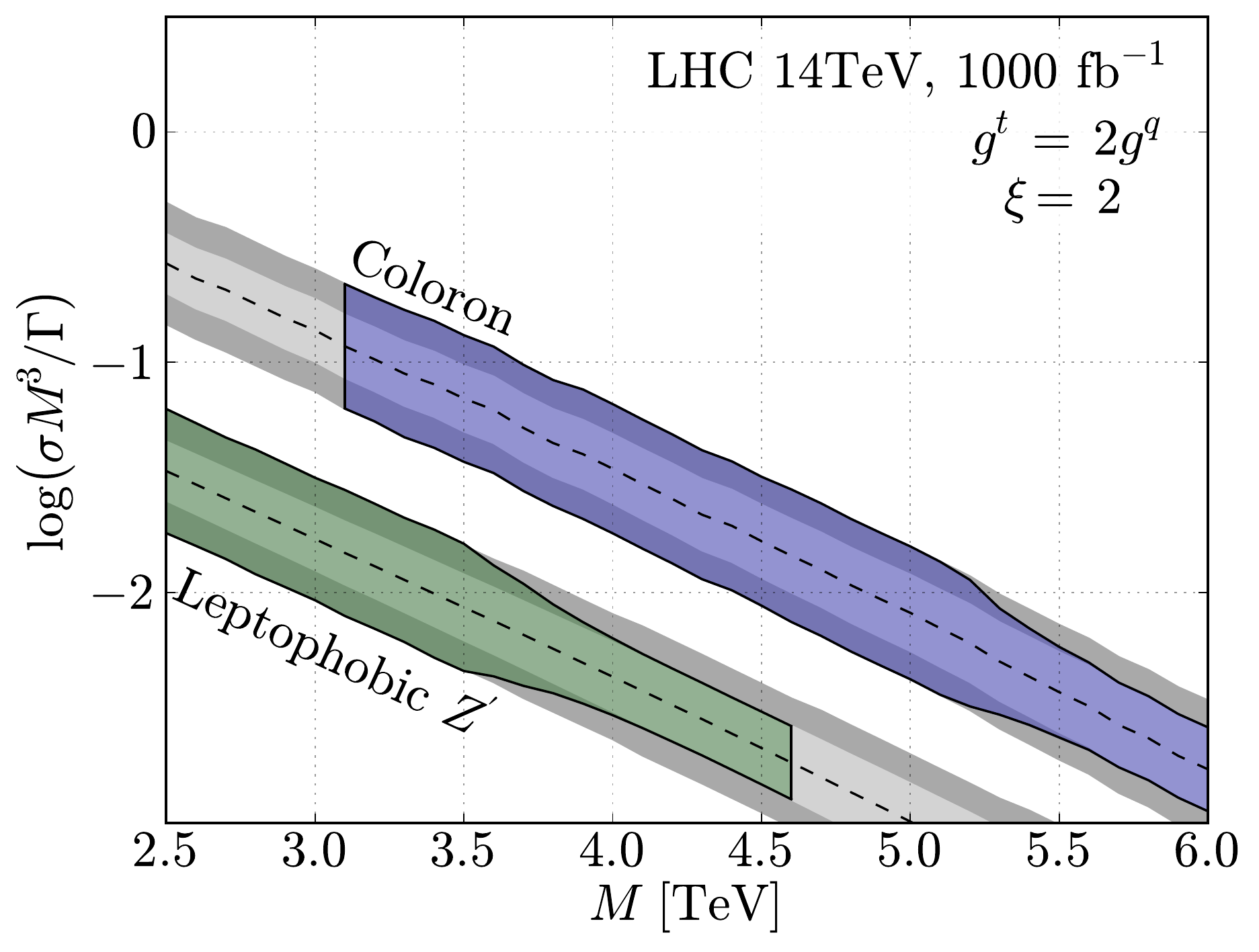}
\includegraphics[width=0.495\textwidth, clip=true]{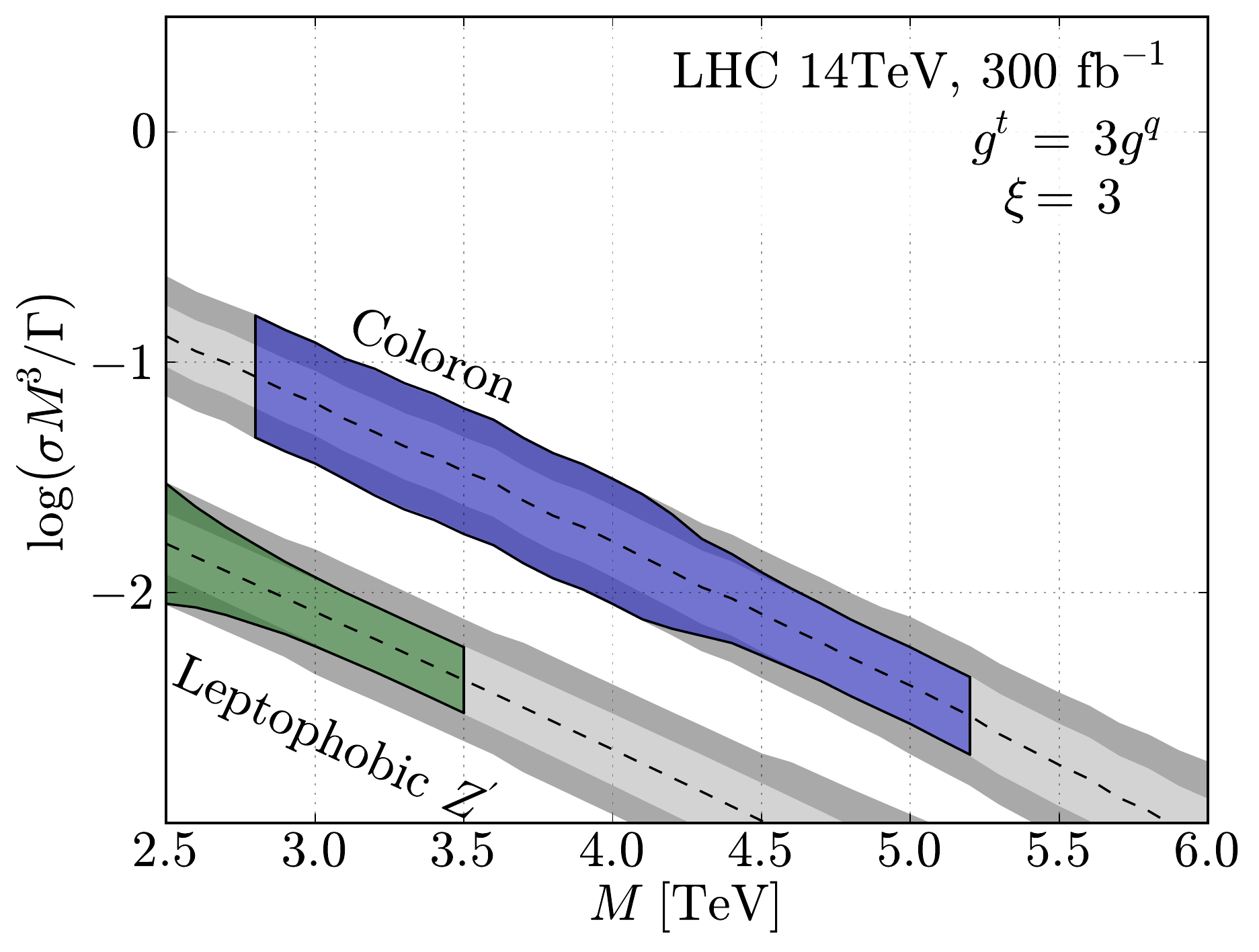}
\includegraphics[width=0.495\textwidth, clip=true]{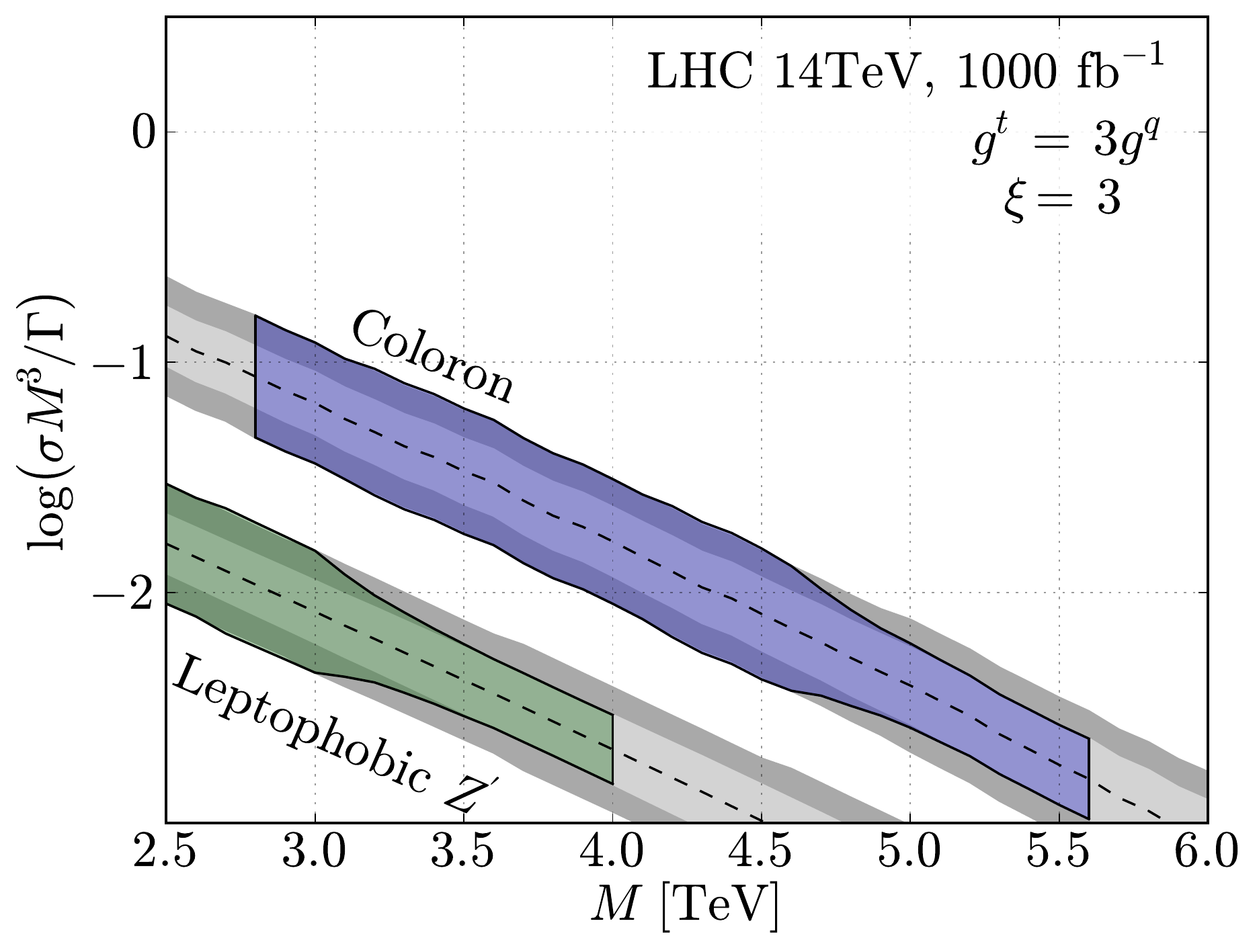}}
\caption{(a) Top left: same as Fig.~\ref{fig:sensuniv}(c), but for the flavor non-universal case with $g_{L,R}^t = 2g_{L,R}^q$. (b) Top right: same as (a) but for integrated luminosity of $1000\ \ifb$. (c) Bottom left: same as (a) but for $g_{L,R}^t = 3g_{L,R}^q$. (d) Bottom right: same as (c) but for integrated luminosity of $1000\ \ifb$.}
\label{fig:sensnon}
\end{figure}

The results for the flavor non-universal case in Fig.~\ref{fig:sensnon} illustrate several features. As discussed in Sec.~\ref{sec:coldis}, the ratio between the widths of a color-octet and color-singlet depends only on the color factors and hence the color discriminant variable can be used to distinguish color-octet and color-singlet resonances in the flavor non-universal scenario for a given $\xi$. This is demonstrated by the separation between the bands for colorons and leptophobic $\zp$s in Fig.~\ref{fig:sensnon}, for the cases $\xi=2$ and $\xi=3$. Several features present in the flavor universal case are preserved. The reach in mass increases with increasing luminosity; the uncertainty in the estimation of $\dcol$ is large for larger masses and is dominated by systematic uncertainties when the number of events is large and the uncertainties for resonances with small (large) widths is large (small). There are also some new features specific to the flavor non-universal case. The reach in mass decreases with increasing $\xi$. As discussed in Sec.~\ref{sec:constr}, as $\xi$ changes the production cross section remains the same while the branching fraction to jets decreases leading to a smaller di-jet cross section and hence fewer signal events. Note that the central value of $\dcol$ decreases with increasing $\xi$ due to the dependance of the di-jet cross section on $\xi^2$ as shown in Eqs.~(\ref{eq:colxsecnon}) and (\ref{eq:zpxsecnon}).

So far we have presented results for distinguishing a color-octet and a color-singlet state in the flavor universal case and in the flavor non-universal case for a given $\xi$. In Fig.~\ref{fig:xilog} we present the sensitivity of the color discriminant variable for varying $\xi$ at the LHC with $\sqrt{s} = 14\ \tev$. In Fig.~\ref{fig:xilog}(a) and (b) we present results for $M=3\ \tev$ and integrated luminosities of $300\ \ifb$ and $1000\ \ifb$ respectively while Fig.~\ref{fig:xilog}(c) and (d) are for $M=4\ \tev$ and integrated luminosities of $300\ \ifb$ and $1000\ \ifb$ respectively. As before the central values are indicated by dashed (black) lines and the reach for colorons (leptophobic $\zp$s) are denoted by blue (green) regions. The different points marked $a,b,c,d$ correspond to different parameter points used as examples below.

\begin{figure}[t]
{\includegraphics[width=0.495\textwidth, clip=true]{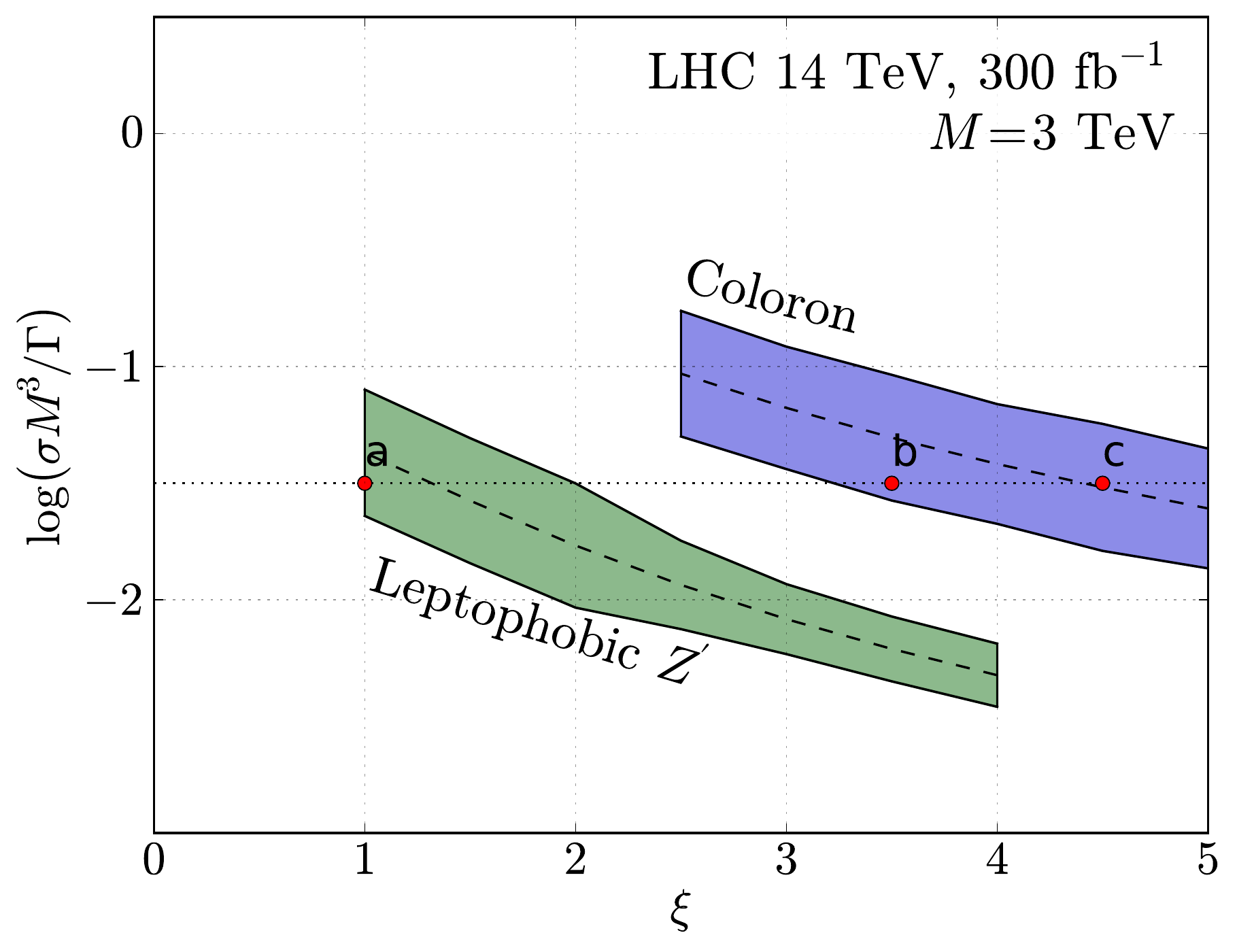}
\includegraphics[width=0.495\textwidth, clip=true]{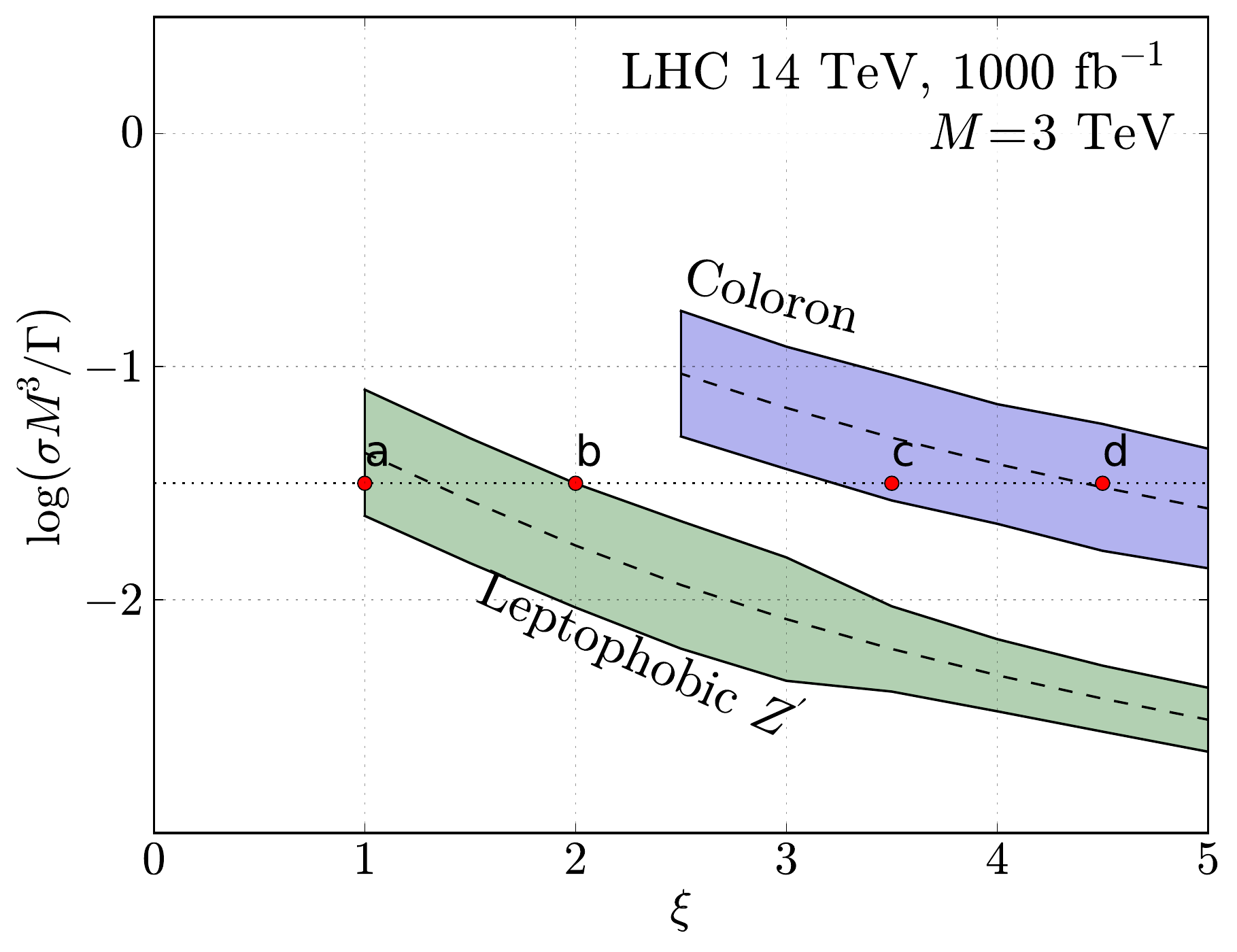}
\includegraphics[width=0.495\textwidth, clip=true]{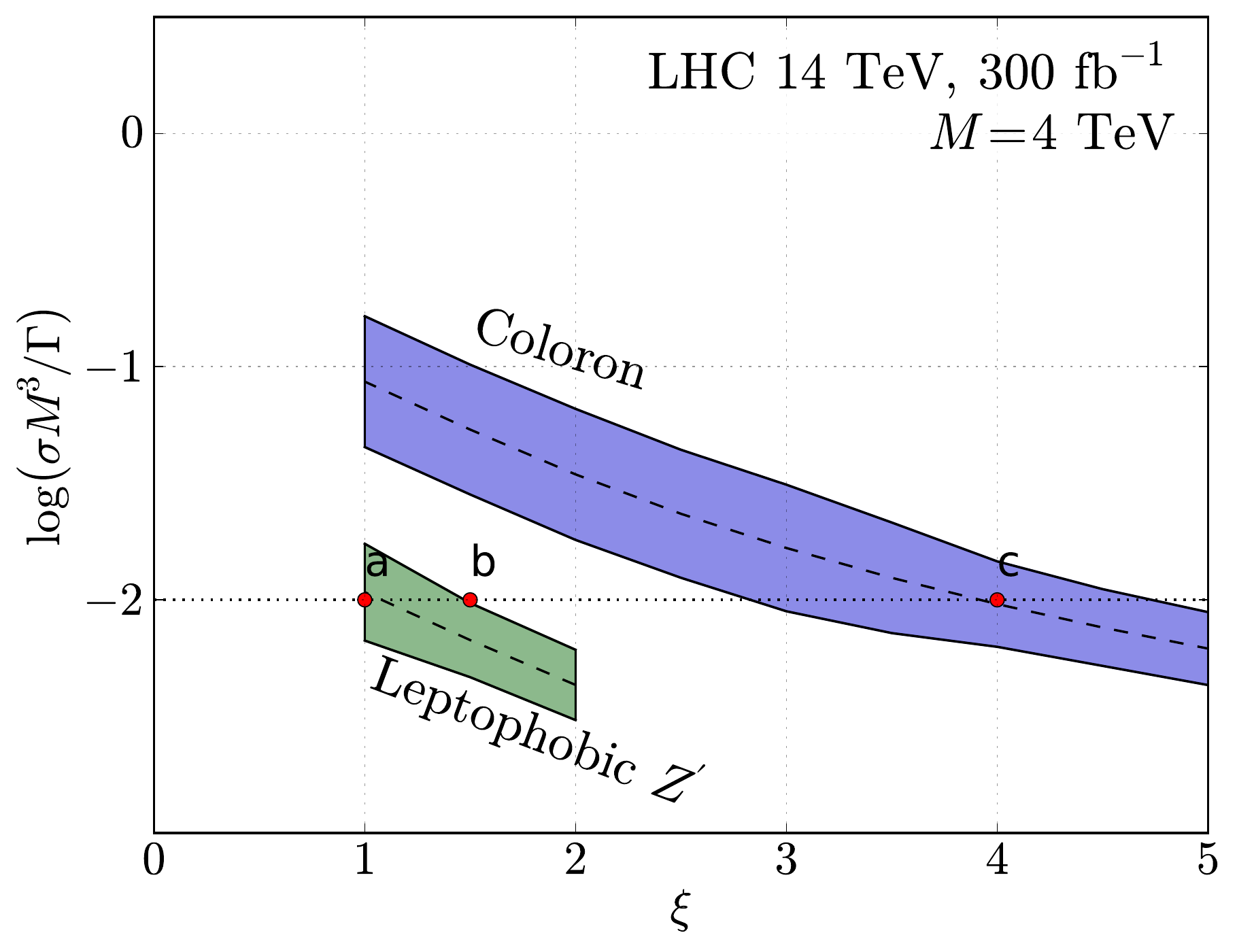}
\includegraphics[width=0.495\textwidth, clip=true]{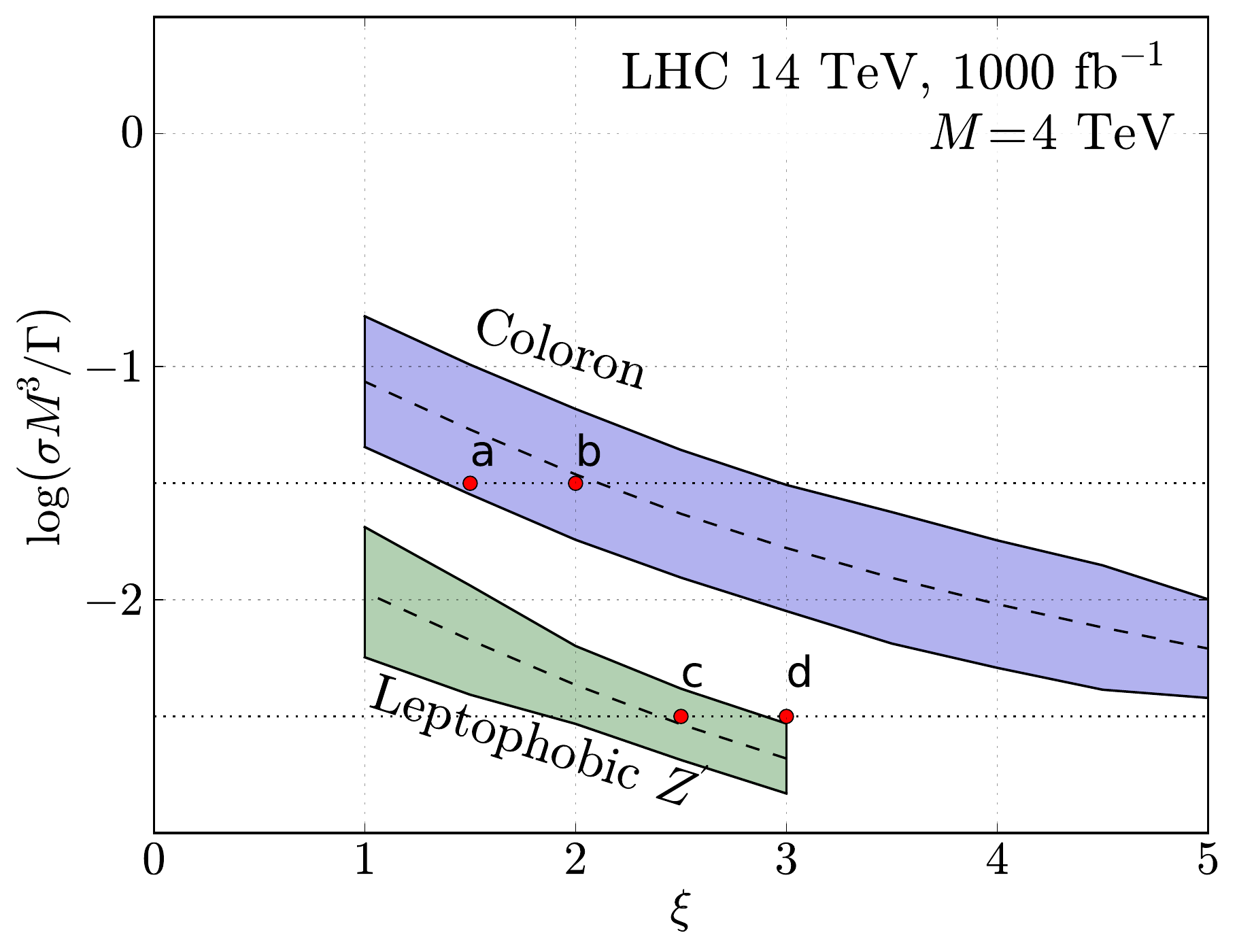}}
\caption{(a) Top left: Sensitivity to distinguish color-octet and color-singlet scenarios at the LHC with $\sqrt{s}=14\ \tev$ and integrated luminosity of $300\ \ifb$ for a resonance of mass $3\ \tev$ for different values of $\xi$. (b) Top right: same as (a) but for integrated luminosity of $1000\ \ifb$. (c) Bottom left: same as (a) but for a resonance with mass $4\ \tev$. (d) Bottom right: same as (c) but for integrated luminosity of $1000\ \ifb$. The different points marked $a,b,c,d$ correspond to different parameter points used as examples in the text.}
\label{fig:xilog}
\end{figure}

With the discovery of a resonance in the di-jet channel a measurement of the mass, width and cross section and hence of the color discriminant variable is possible. Several scenarios may be allowed for a given measured value of mass and $\dcol$. For example, in Fig.~\ref{fig:xilog}(a), for ${\rm{log}}(\dcol)=-1.5$  a few of the allowed possibilities include a leptophobic $\zp$ with $\xi=1$ and a coloron with $\xi=3.5\ \rm{or}\  4.5$ and are marked by the points labeled $a,b,c$ respectively. Similar possibilities are shown for other masses and luminosities in Fig.~\ref{fig:xilog}(b) - (d) for different measured values of $\dcol$ and are labeled as $a,b,c,d$. With just the measurements from the di-jet discovery channel it would not be possible to distinguish which of these scenarios is being realized in nature. Additional information is required, for instance, by measuring $\xi$ from the ratio of the cross sections when the resonance decays to top pairs vs jets as given in Eq.~(\ref{eq:xi}). With these two measurements, {\it i.e.} $\dcol$ and $\xi$, one can distinguish between color-octet and color-singlet resonances and their flavor structure. 

Finally, we note that distinguishing between color-octet and color-singlet resonances of various flavor structures depends on the reach of $\xi$ as well as the precision with which $\xi$ can be measured at the LHC. This in turn depends on the measurements of the cross section in the di-jet channel and the top pair channel and the associated uncertainties. For example, a resonance may be discovered at a given mass in the di-jet channel and a measurement of $\dcol$ may be made. However a measurement of the cross section in the top pair channel may not be possible or may have large uncertainties due to limited statistics (because of small branching to top pairs). In such a case a measurement of $\xi$ is not possible or has very large errors and the ability to distinguish various flavor scenarios is reduced. A detailed exploration of the reach and precision with which $\xi$ can be measured is beyond the scope and focus of this article. Instead we refer the interested reader to experimental studies of resonance searches in the top pair \cite{ATLAS:2009hdz, CMS:2009fwa, CMS:2009cxa, ATLAS:2011gmi} and di-jet channels \cite{Gumus:2006mxa} and the results of such studies can provide an estimate of $\xi$ and the uncertainties in measuring $\xi$. 

\section{Summary}
\label{sec:summ}

Di-jet resonance searches are simple but powerful model-independent probes for discovering new particles motivated in many new physics scenarios. Once a resonance has been discovered in the di-jet channel it will be very important to measure its properties. The di-jet (discovery) channel can provide information about the mass and spin of the resonance as well as constrain the coupling strength. It does not provide information about the chiral structure; that can be obtained by including information from the channel where the resonance is produced in association with a SM electroweak gauge boson. The next question that remains to be resolved is the color structure of the new resonance; in particular, whether it is a color-octet or color-singlet resonance. 

In this article we proposed a new variable called the color discriminant variable ($\dcol$) to distinguish color-octet from color-singlet resonances. This variable is relatively model-independent and is sensitive to the color and flavor structure. In order to make our study as widely applicable as possible, we studied phenomenological models of color-octet and color-singlet resonances without being tied down to a specific theory. As illustrative examples we studied colorons and leptophobic $\zp$s in the flavor universal case as well as an illustrative flavor non-universal scenario. We analyzed the current constraints on the parameter space and the discovery potential of coloron and leptophobic $\zp$ models at the future LHC. We sampled a wide range of masses, couplings and values of $\xi$, which parameterizes the relative coupling strength of the third generation and first generation SM quarks to the new resonance. 

We studied the sensitivity to distinguish color-octet and color-singlet resonances in the flavor universal as well as flavor non-universal scenarios and presented our results for the LHC with c.m. energy of $14\ \tev$ for varying luminosities, ${\cal L} = 30,\ 100,\ 300\ \rm{and}\ 1000\ \ifb$, after including all uncertainties. We  found that our method has a wide reach in mass and couplings as well as $\xi$. A measurement of $\dcol$ alone can distinguish between color-octet and color-singlet states for a given flavor scenario. However to distinguish between different flavor scenarios we will need additional information which comes from a measurement of $\xi$. Together, $\dcol$ and $\xi$ can help distinguish a color-octet from a color-singlet state as well as establish the nature of the couplings in the flavor sector. We find that the LHC will be able to provide information about the color and flavor structure of a new di-jet resonance for a wide range of couplings and masses and hence point us in the direction of the underlying theoretical structure. 

\begin{acknowledgments}

We would like to thank Georges Azuelos, Joey Huston, Jim Linneman and Wade Fisher for useful discussions. This work is supported by the United States National Science Foundation under grant PHY-0854889. We wish to acknowledge the support of the Michigan State University High Performance Computing Center and the Institute for Cyber Enabled Research. PI is supported by Development and Promotion of Science and Technology Talents Project (DPST), Thailand. EHS and RSC thank the Aspen Center for Physics and the NSF Grant No. 1066293 for hospitality during the writing of this paper.
\end{acknowledgments}

\appendix

\section{Uncertainty of Intrinsic Width Measurement}
\label{sec:app}

The uncertainty in the intrinsic width of the resonance plays a key role in the estimation of the uncertainties in $\dcol$, the variable we propose to distinguish a color-octet and a color-singlet state. In this section we extract the uncertainty of the intrinsic width from a measurement of the total width and a knowledge of the systematic uncertainties in measuring that width. The systematic uncertainties relevant to width measurement are the di-jet mass resolution and the uncertainty in the di-jet mass resolution; we model them to have a Gaussian distribution and ignore correlations between them.

The standard deviation of the observed invariant mass distribution ($\sigma_T$) is related to the standard deviation of the intrinsic width ($\sigma_\Gamma \simeq \Gamma/2.35$ assuming a Gaussian distribution) and that of the detector mass resolution ($\mres$) by
\beq
\sigma_\Gamma = \sqrt{ \sigma_T^2 - \mres ^2 } \,.
\label{eq:sigtot}
\eeq
This implies that
\beq
\left(\Delta\sigma_\Gamma\right)^2 =   \left( \frac{\sigma_T \Delta\sigma_T}{\sqrt{\sigma_T^2 - \mres^2} } \right)^2 
+  \left( \frac{\mres \Delta \mres }{\sqrt{\sigma_T^2 - \mres ^2} } \right)^2  \,,
\eeq
or
\beq
\left( \frac{\Delta\sigma_\Gamma}{\sigma_\Gamma}\right)^2 =   \left(1 + \frac{\mres ^2}{\sigma_\Gamma^2}\right)^2
\left( \frac{\Delta\sigma_T}{\sigma_T} \right)^2 
+  \left( \frac{\mres ^2}{\sigma_\Gamma^2}\right)^2  \left( \frac{\Delta \mres }{ \mres } \right)^2   \,,
\label{eq:delta-sigma}
\eeq
where Eq.~(\ref{eq:sigtot}) was used to obtain Eq.~(\ref{eq:delta-sigma}). 

For $N$ observed  signal events, where $N$ is sufficiently large, the uncertainty of the observed width $\Delta\sigma_T $ is given by $\sigma_T/\sqrt{2(N-1)}$ \cite{Beringer:1900zz}.  So Eq.~(\ref{eq:delta-sigma}) leads to
\beq
\frac{\Delta\sigma_\Gamma}{\sigma_\Gamma}=   \sqrt{  \left[1 + \left(\frac{\mres}{\sigma_\Gamma}\right)^2 \right]^2
\frac{1}{2(N-1)}
+  \left(\frac{\mres}{\sigma_\Gamma}\right)^4  \left( \frac{\Delta \mres}{\mres} \right)^2  } \,,
\eeq
where $\Delta\sigma_\Gamma/\sigma_\Gamma = \Delta\Gamma/\Gamma$. Note that for large $N$, the above expression simplifies to
\beq
\frac{\Delta\Gamma}{\Gamma} =  \left( \frac{\mres}{\sigma_\Gamma}\right)^2  \left( \frac{\Delta \mres}{\mres} \right)   \, = 
\left( \frac{\mres}{\Gamma/2.35}\right)^2  \left( \frac{\Delta \mres}{\mres} \right)   \,. 
\eeq
This expression shows that the uncertainty in the intrinsic width is inversely proportional to $\Gamma/\mres$ which implies that the uncertainty in the intrinsic width is small (large) when the intrinsic width is large (small).

\bibliography{}

\end{document} 
